\documentclass[%
 reprint,
superscriptaddress,
 amsmath,amssymb,
 aps,
prl,
floatfix,
nofootinbib,
]{revtex4-2}

\usepackage{graphicx}
\usepackage{dcolumn}
\usepackage{bm}

\usepackage{orcidlink}
\usepackage{xcolor}

\usepackage{booktabs}
\usepackage{accents}
\usepackage{threeparttable}

\newcommand{\dbtilde}[1]{\accentset{\approx}{#1}}

\begin{document}

\title{Nuclear moments, charge radii, and magnetization distribution parameters of Ag isotopes from laser spectroscopy and \textit{ab initio} electronic-structure calculations}

\author{Leonid V.\ Skripnikov}
\email{skripnikov\_lv@pnpi.nrcki.ru,\\ leonidos239@gmail.com}
\homepage{http://www.qchem.pnpi.spb.ru}
\affiliation{Petersburg Nuclear Physics Institute named by B.P. Konstantinov of National Research Centre
``Kurchatov Institute'', Gatchina, Leningrad District 188300, Russia}
\affiliation{Saint Petersburg State University, 7/9 Universitetskaya nab., St. Petersburg, 199034 Russia}

\author{Bram van den Borne}
\email{bram.petrus.van.den.borne@cern.ch}
\altaffiliation[Present address: ]{CERN, CH-1211 Geneva, Switzerland}
\affiliation{KU Leuven, Instituut voor Kern- en Stralingsfysica, B-3001 Leuven, Belgium}

\author{Michail Athanasakis-Kaklamanakis}
\email{m.athkak@cern.ch}
\altaffiliation[Present address: ]{JILA and University of Colorado, Boulder, Colorado 80309, USA}
\affiliation{Centre for Cold Matter, Imperial College London, SW7 2AZ London, United Kingdom}

\author{Gleb Penyazkov}
\email{penyazkovg10@mails.tsinghua.edu.cn}
\affiliation{State Key Laboratory of Low-Dimensional Quantum Physics, Department of Physics, Tsinghua University, Beijing 100084, China}

\author{Ruben P. de Groote}
\email{ruben.degroote@kuleuven.be}
\affiliation{KU Leuven, Instituut voor Kern- en Stralingsfysica, B-3001 Leuven, Belgium}

\date{September 30, 2026}

\begin{abstract}
The nuclear electromagnetic moments and mean-square charge radii of several silver (Ag) isotopes deduced from recent laser spectroscopy studies in the mass region $A=$ 96-121 are determined using high-accuracy electronic structure calculations performed in this work. We report hyperfine structure and isotope-shift atomic factors calculated with the relativistic coupled cluster approach, including single, double, triple, and perturbative quadruple excitations, CCSDT(Q). Following a systematic theoretical uncertainty analysis, we show that at the precision now achieved (sub-percent uncertainties in some cases), quantum electrodynamic effects become significant. We also show that the isotope-dependent effect in the hyperfine structure due to the non-point-like nuclear magnetization distribution can be extracted with negligible dependence on the assumed nuclear magnetization model at the present level of precision. This also yields a nuclear magnetic dipole moment that is corrected for the hyperfine anomaly induced by the Bohr-Weisskopf effect. In terms of the nuclear electric quadrupole moments, the uncertainty in the electric-field gradient used to extract the quadrupole moments from laser spectroscopy has also been reduced by one to two orders of magnitude relative to values used in previous studies. Finally, the nuclear charge radii of Ag isotopes are extracted using field- and mass-shift factors from our coupled cluster calculations, and the difference in mean-square charge radii between $^{107,109}$Ag agrees with the value deduced from muonic X-ray spectroscopy.
\end{abstract}

\maketitle

\section{Introduction} \label{sec:intro}
With three protons below the $Z=50$ proton shell closure, the isotopic chain of Ag ($Z=47$), which extends from below the neutron shell closure at $N=50$ to above the one at $N=82$, is a sensitive probe of nuclear structure near doubly magic nuclei. The wider interest in the structure of nuclei around $Z=50$ has been strengthened in recent years by developments in \textit{ab initio} nuclear theory~\cite{Ekstrom2023AbInitio}, which has enabled calculations in medium-mass nuclei using nucleon-nucleon interactions derived from the underlying quantum chromodynamics using effective field theory treatments.

Nuclear magnetic dipole and electric quadrupole moments provide information on the nuclear wavefunction, on the deviation of nuclear shapes from sphericity, and on the interplay of single-particle behavior and collectivity~\cite{Neyens2003}. The evolution of nuclear charge radii across an isotopic chain, on the other hand, provides insights into complex nuclear phenomena such as shape coexistence, regions of nuclear deformation, the presence of halo-like nucleon orbitals, and more~\cite{NorterhauserMooreRadii2020}. Measurements of nuclear charge radii and electromagnetic moments have become important benchmarks for state-of-the-art calculations of medium-mass nuclei. Recently, numerous studies using nuclear density functional theory, valence-space in-medium similarity renormalization group, and large-scale shell model calculations have focused on the structure of Ag isotopes and their vicinity~\cite{ferrer2014,degroote2024AgPLB,Nies2023,Ge2024,Geldhof2022,Reponen2021,vandenborne2025,Yordanov2020,Maas2025Ru,Sassarini2022DFT}.

Beyond its relation to the $Z=50$, $N=50,82$ shell closures, the Ag isotopic chain is distinguished by the presence of isomeric nuclear states for nearly every isotope. This feature of Ag offers the opportunity to trace the evolution of multiple nuclear states with distinct configurations and different spins as a function of neutron number, which is possible for both odd-odd and odd-even isotopes, unlike the well-studied even-$Z$ isotopic chains.

\begin{figure*}[htb]
    \centering
    \includegraphics[width=0.7\linewidth]{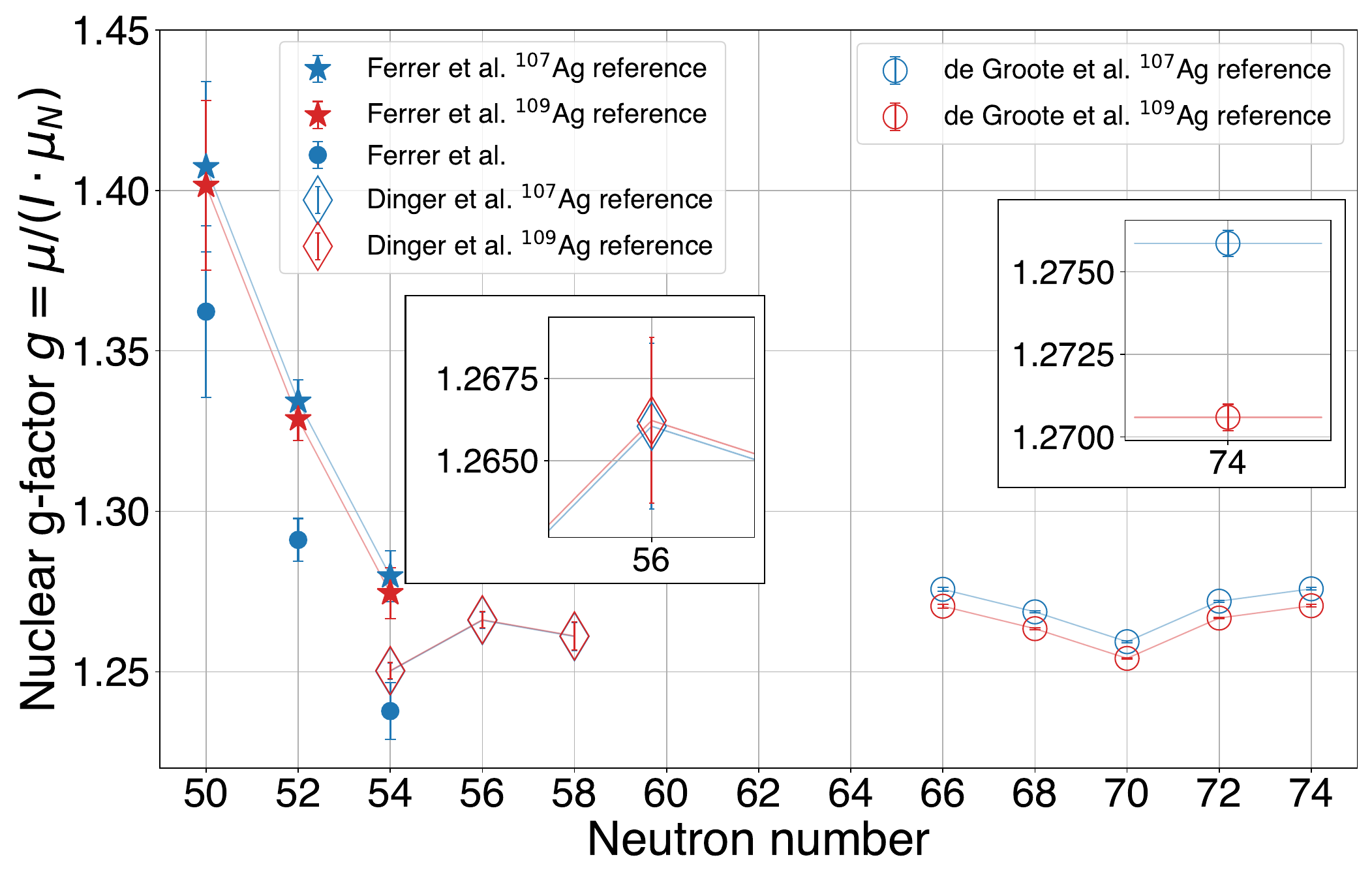}
    \caption{Overview of nuclear $g$-factors of odd-even Ag isotopes with spin $I = 7/2^+$ ($I=9/2^+$ for $^{97,99,101}$Ag) from laser spectroscopy. Error bars show the experimental statistical uncertainty. The uncertainties originating from the reference moments are given by the colored band, visible as a line on the figure. Hyperfine parameters reported by Ferrer \textit{et al.}~\cite{ferrer2014}, Dinger \textit{et al.}~\cite{dinger1989} and de Groote \textit{et al.}~\cite{degroote2024AgPLB} were used to extract $g$-factors relative to two reference isotopes. The reference magnetic moments were taken from the recommended values in the compilation by Mertzimekis \cite{mertzimekis2016}. The $A^{\rm{ref}}$-factors of the reference isotopes are taken from Ref.~\cite{dahmen1967} for the ground-state (measured by Ferrer \textit{et al.} and de Groote \textit{et al.}) and from Ref.~\cite{blachman1966} for the state investigated by Dinger \textit{et al.} The data series labeled ``Ferrer \textit{et al.}'' denotes the originally published $g$-factors in Ref.~\cite{ferrer2014}.}
    \label{fig:lit_g-factor}
\end{figure*}

Extracting nuclear moments and radii of radioactive isotopes from laser spectroscopy requires knowledge of the atomic electromagnetic fields they interact with, which are generated by the motion of the atomic electrons. These fields within the Ag atom could not be calculated accurately until recently, and thus nuclear moments have typically been extracted from the measured hyperfine spectra using a precise non-optical measurement of the magnetic dipole moment of one of the stable Ag isotopes as a reference value, as done in Refs.~\cite{fischer1975,dinger1989,ferrer2014,Reponen2021,degroote2024AgPLB,vandenborne2025}. However, different values for these stable reference magnetic moments appear in the literature.
The magnetic dipole moments of the stable $^{107}$Ag and $^{109}$Ag isotopes, which have been studied by nuclear magnetic resonance~\cite{burges1973,sahm1974} and corrected by Raghavan~\cite{Raghavan:89}, appear in the compilation of Mertzimekis~\cite{mertzimekis2016}, while the table of recommended magnetic dipole moments by Stone~\cite{stone2019} differs significantly from these. The most recent recommended magnetic moments for the stable Ag isotopes were presented by Antu\v{s}ek and Repisky~\cite{antusek2020}. However, these values are not yet cited in the evaluated nuclear databases.

Magnetic dipole moments from atomic laser spectroscopy data are furthermore sensitive to a non-negligible correction known as the hyperfine anomaly (HFA), which is a small correction to the overall atomic hyperfine splitting but can become comparable to the experimental uncertainty in high-precision measurements. When extracting the moments of radioactive isotopes relative to that of a reference isotope, significant inconsistencies can appear if the HFA correction is not taken into account. Especially in the Ag isotopes, where the HFA in the $4d^{10}5s~^{2}S_{1/2}$ ground state is large, differential anomalies as large as $-3.4(17)\%$ between the ground states of $^{103,107}$Ag~\cite{wannberg1970,Persson:2023} and $-0.41277(7)\%$ between the ground states of $^{107,109}$Ag~\cite{burges1973} have been reported.

In order to illustrate the effect of the differential HFA between $^{107}$Ag and $^{109}$Ag that is known to exist in the $4d^{10}5s\,^{2}S_{1/2}$ atomic ground state, Fig.~\ref{fig:lit_g-factor} shows the $g$-factors of Ag isotopes extracted from the experimental hyperfine splittings using the recommended magnetic moments from Ref.~\cite{Raghavan:89} for $^{107}$Ag or $^{109}$Ag as the reference value. As de Groote \textit{et al.}~\cite{degroote2024AgPLB} measured the hyperfine structure of the $^2S_{1/2}$ atomic state with high precision, even the small HFA between the two stable isotopes used as reference values leads to $g$-factors that are significantly different. The low-precision values obtained by Ferrer \textit{et al.}~\cite{ferrer2014} are not sensitive to the choice of reference value given their larger statistical error. Note, however, that Ferrer \textit{et al.} had corrected their magnetic moments for the large differential HFA with respect to the $^{109}$Ag reference using the semi-empirical approach of Moskowitz and Lombardi~\cite{moskowitz1973}, which led to good agreement of their $^{101}$Ag value with that of Dinger \textit{et al.}~\cite{dinger1989}. The latter was measured in the $4d^95s^2\,^2D_{5/2}$ state, which is not affected by the HFA within the level of achieved experimental precision.

In order to extract accurate HFA-corrected moments and radii from high-precision laser spectroscopy data, we have performed high-accuracy \textit{ab initio} electronic-structure calculations for the hyperfine and isotope-shift factors of the $4d^{10}5s\,^2S_{1/2}$, $4d^{10}5p\,^2P_{3/2}$, $4d^{10}5p\,^2P_{1/2}$, and $4d^95s^2\,^2D_{5/2}$ states in neutral Ag, as well as the NMR shielding constant relevant to literature NMR measurements with Ag(H$_2$O)$^+_4$. Combining these results with laser-spectroscopy data reported in Refs.~\cite{ferrer2014,dinger1989,degroote2024AgPLB,vandenborne2025}, we present the resulting nuclear magnetic dipole moments, electric quadrupole moments, and charge radii from $^{96}$Ag to $^{121}$Ag, which address several past inconsistencies. We demonstrate how the HFA-corrected nuclear magnetic dipole moment and the nuclear magnetization distribution parameter can be determined for a wide range of Ag isotopes with negligible dependence on the assumed nuclear magnetization model at the present level of precision, using the measured hyperfine splittings of two or more atomic states. 

This paper is structured into three parts, discussing in turn the magnetic dipole moment, the electric quadrupole moment, and the changes in the nuclear charge radii. Within these three parts, we organize the discussion along the following subsections: for magnetic dipole moments, we first discuss the interpretation of nuclear magnetic resonance (NMR) through the lens of new calculations. This is an essential first step to ensure the accuracy of the nuclear magnetic moments. We then proceed with a discussion of the calculated atomic hyperfine fields required to extract the nuclear moments and to study the Bohr-Weisskopf effect in the stable isotopes. This provides the basis for an investigation of the influence of the Bohr-Weisskopf effect in the radioactive isotopes of silver as well. For the nuclear quadrupole moments, we present new calculations of the atomic field gradients, and use them to extract the moments for all isotopes and nuclear states of Ag where atomic hyperfine splittings have been used to determine the quadrupole moment. A similar approach is used for the nuclear charge radii: we start by presenting new calculations of the atomic factors, and then compare the extracted charge radii with muonic spectroscopy and previous optical determinations of nuclear charge radii.

\section{Nuclear magnetic dipole moments} \label{sec:results}
\subsection{NMR shielding}
The NMR shielding constant quantifies the screening of the externally applied magnetic field experienced by the nucleus under study due to the local chemical environment. As a result, taking into consideration the NMR shielding constant is necessary for the extraction of accurate values of the nuclear magnetic dipole moment from NMR experiments. We begin by calculating this shielding constant and comparing it to the available literature.

One can define the shielding tensor associated with nucleus $j$ in a given molecule using the following expression:
\begin{equation}
\label{SHIELDINGDer}
\left.\sigma^j_{a,b}=\frac{\partial^2E}{\partial\mu_{j,a}\partial {\rm B}_b} \right|_{{\bm{\mu}_j=0,{\rm \bf{B}}=0}}.
\end{equation}
Here, $E$ is the system's energy, $\mu_{j,a}$ denotes the $a$'th component of the nuclear magnetic moment vector $\bm{\mu}_j$ for nucleus $j$, and ${\rm B}_b$ stands for the $b$'th component of the uniform external magnetic field vector ${\rm \bf{B}}$. To interpret molecular NMR experiments conducted in a solution, we require the isotropic part $\sigma$ of the shielding tensor, calculated as $\sigma= \frac{1}{3}\sum_a\sigma_{a,a}$. In an NMR experiment, one obtains the uncorrected value $\mu^{\rm uncorr}$ for the nuclear magnetic dipole moment, i.e., the value not adjusted for magnetic shielding effects. The intrinsic magnetic moment $\mu$ can then be extracted as:

\begin{equation}
\label{mumuuncorr}
\mu=\mu^{\rm uncorr} / (1-\sigma).
\end{equation}

The effect of the external uniform magnetic field \textbf{B} on electrons within a molecule is described by the following term in the Dirac-Coulomb Hamiltonian:
\begin{equation}
 \label{HB}
{\rm H}_B={\rm \bf{B}}\cdot \frac{1}{2}(\bm{r}_G \times \bm{\alpha}),
\end{equation}
where $\bm{\alpha}$ represents Dirac matrices and $\bm{r}_G = \bm{r} - \bm{R}_G$, with $\bm{R}_G$ being the gauge origin~\cite{dyall2007}. This gauge origin is the reference point for the coordinate system used to describe the electron radius vector in this equation. The choice of $\bm{R}_G$ can potentially affect the numerical result for the shielding constants, especially when employing modest basis sets. This uncertainty, however, can be controlled by considering the convergence with respect to the increasing basis-set.

In the point-magnetic-dipole approximation, the interaction between an electron and the magnetic moment $\bm{\mu}_j$ of the $j$th nucleus is given by:
\begin{equation}
 \label{HHFS}
{\rm H}_{\rm hyp}= \bm{\mu}_j\cdot \frac{(\bm{r}_j \times \bm{\alpha})}{r_j^3},
\end{equation}
where $\bm{r}_j=\bm{r} - \bm{R}_j$, and $\bm{R}_j$ denotes the position of nucleus $j$.

For a single-particle case, the tensor (Eq.~\eqref{SHIELDINGDer}) can be computed using the sum-over-states method in second-order perturbation theory with perturbations (Eqs.~\eqref{HB}, \eqref{HHFS}):
\begin{eqnarray}
\label{SHIELDPT}
& & \sigma_{a,b}=\\
& & \sum_{n \neq 0}
\frac{\langle 0 | (\frac{(\bm{r}_j \times \bm{\alpha})}{r_j^3})_a |n \rangle
\langle n| (\frac{1}{2}(\bm{r}_G \times \bm{\alpha}))_b |0 \rangle}
{E_0-E_n} + \text{h.c.} \nonumber,
\end{eqnarray}
where $|0 \rangle$ represents the unperturbed one-particle state, $|n \rangle$ is the unoccupied $n$'th unperturbed state (orbital), and h.c. denotes the Hermitian conjugate. Within the Dirac theory, the summation should encompass both positive energy and negative energy states $|n \rangle$~\cite{aucar1999}. The component associated with positive energy states is termed the ``paramagnetic'' part, while that with negative energy states is termed the ``diamagnetic'' part~\cite{aucar1999}. For Dirac-Hartree-Fock (DHF) and density functional theory (DFT) many-electron methods, the response technique can be employed to compute both terms~\cite{olejniczak2012,ilias2013,aucar1999,dirac19}. A more accurate result for the paramagnetic part of the shielding constant, with clearer control of the uncertainty, can be achieved by using relativistic coupled cluster theory~\cite{skripnikov2018}. We follow this approach in the present paper.

\subsection{Reference magnetic dipole moments of stable isotopes from NMR}\label{subsec:NMR}

Details on the computational methods for the calculation of all properties in this work are given in the Appendix. Table~\ref{ShieldCalc} gives the calculated contributions to the shielding constant for the NMR measurements in water~\cite{sahm1974}. The dominant contribution to the shielding constant arises from the diamagnetic part. Using relativistic CC theory enables systematic control over the convergence of results related to the electron correlation treatment of the paramagnetic part. This control is achieved by comparing the results of CCSD(T) to those obtained using the CCSD approach. The difference between these methods is small and indicated as ``corr'' in Table~\ref{ShieldCalc}. The largest source of uncertainty originates from the solvent effect. The most detailed estimation of this effect and its associated uncertainty, 60 ppm, was provided in Ref.~\cite{antusek2020} and is included in Table \ref{ShieldCalc}. The value for the basis-set correction is used as a measure of the uncertainty due to basis set incompleteness. Additionally, there is an uncertainty due to the geometric parameters indicated as ``geom'' in Table~\ref{ShieldCalc}. This uncertainty is determined by comparing shielding constant values calculated at the relativistic PBE0~\cite{pbe0} level using two sets of Ag(H$_2$O)$_4^+$ geometric parameters: one optimized at the relativistic PBE0~\cite{pbe0} level and the other based on the experimental geometry of Ref.~\cite{persson2006coordination}.

The final value of $\sigma = 4496(76)$~ppm of the shielding constant, obtained by combining the relativistic CC and DFT theory methods, shows good agreement with the value of $\sigma = 4483(61)$~ppm reported in Ref.~\cite{antusek2020}, where a combination of non-relativistic CC and relativistic DFT methods was used to determine $\sigma$. The slightly larger uncertainty in the present work stems from different uncertainty estimation schemes and the inclusion of more uncertainty sources.

\begin{table}[h]
\caption{Calculated values (in ppm) of the contributions to the Ag shielding constant $\sigma$ for Ag(H$_2$O)$_4^+$.} 
\label{ShieldCalc}
\begin{tabular}{lr}
\hline
\hline
Contribution &  Value \\
\hline

Diamagnetic:                                           \\
\hspace{0.5cm}PBE0                         & 4365$(1)_{\rm geom}$                  \\
Paramagnetic:                                          \\
\hspace{0.5cm}CCSD(T)~~                    & 374$(8)_{\rm corr}(38)_{\rm geom}$                  \\
\hspace{0.5cm}Basis-set correction         & $-24(24)$                 \\
Gaunt                          & $-5(5)$                  \\
Solvent effect, from Ref.~\cite{antusek2020} & $-214(60)$   \\
\\
Total                  & 4496(76) \\
\hline
\hline
\end{tabular}
\end{table}

\begin{table}[h]
\caption{\label{tab:NMR_comparison} Comparison of the originally reported and re-evaluated magnetic dipole moments of $^{107,109}$Ag from NMR measurements of Ag in water.
}
\begin{tabular}{lll}
\hline
\hline
                            & \hspace{0.4cm}$\mu$ / $\mu_N$           & \hspace{0.4cm}Reference \\ \hline
$^{107}$Ag &                  \hspace{0.4cm}$-0.1130455(7)$           & \hspace{0.4cm}Sahm and Schwenk~\cite{sahm1974} \\
$^{109}$Ag &                  \hspace{0.4cm}$-0.1299615(8)$           & \hspace{0.4cm}Sahm and Schwenk~\cite{sahm1974}   \\
 & & \\
$^{107}$Ag &                  \hspace{0.4cm}$-0.113555(7)$            & \hspace{0.4cm}Antu\v{s}ek and Repisky~\cite{antusek2020} \\
$^{109}$Ag &                  \hspace{0.4cm}$-0.130555(8)$            & \hspace{0.4cm}Antu\v{s}ek and Repisky~\cite{antusek2020} \\
 & & \\
$^{107}$Ag &                  \hspace{0.4cm}$-0.113558(9)$            & \hspace{0.4cm}This work \\
$^{109}$Ag &                  \hspace{0.4cm}$-0.130551(10)$           & \hspace{0.4cm}This work \\
\hline
\hline
\end{tabular}
\end{table}

The ratios of the Larmor frequencies $\nu(^{107}{\rm Ag})/\nu(^{1}{\rm H})$ and $\nu(^{109}{\rm Ag})/\nu(^{1}{\rm H})$ were measured in Ref.~\cite{sahm1974}. By substituting our calculated value of $\sigma_{\rm Ag}$ for Ag and the shielding constant $\sigma_{^1 {\rm H}}$ for $^1$H in water~\cite{Garbacz:2019,Gottlieb1997} together with the most recent value of the bare proton magnetic moment~$\mu_p=~2.792\:847\:344\:62(82)\,\mu_\mathrm{N}$~\cite{Schneider2017} into the equation,
\begin{equation}
 \mu_{\rm Ag} = \frac{I_{\rm Ag}}{I_{\rm H}} \frac{\nu_{\rm Ag}}{\nu_{^1 {\rm H}}}\frac{1-\sigma_{^1 {\rm H}}}{1-\sigma_{\rm Ag}}\,\mu_{\rm p}.
\end{equation}
one obtains
\begin{eqnarray}
\label{mu107}
\mu({\rm ^{107}Ag})=-0.113558(9) \mu_N, \\    
\label{mu109}
\mu({\rm ^{109}Ag})=-0.130551(10) \mu_N.
\end{eqnarray}

Table~\ref{tab:NMR_comparison} shows the values of the uncorrected magnetic dipole moments of $^{107}$Ag and $^{109}$Ag as originally reported by Sahm and Schwenk~\cite{sahm1974}, and the values from the re-evaluation of the NMR data by Antu\v{s}ek and Repisky~\cite{antusek2020} and this work, both estimating the shielding constant using modern theoretical approaches. The two sets of corrected moments are in very good agreement. The corrected values in the present work confirm the results by Antu\v{s}ek and Repisky~\cite{antusek2020} and should not be seen as superseding them, but the remainder of this work uses the present values of $\mu$ for $^{107,109}$Ag, rather than those by Antu\v{s}ek and Repisky.

As the shielding correction does not change the ratio $\mu(^{107}{\rm Ag})/\mu(^{109}{\rm Ag})$, using these new reference moments to calculate the nuclear $g$-factors of radioactive isotopes shown in Fig.~\ref{fig:lit_g-factor} does not resolve the discrepancy between the use of the two reference isotopes for magnetic moments measured in the $5s\,^2S_{1/2}$ atomic level. To resolve this, the BW effect should be included.

\subsection{Magnetic dipole hyperfine interaction theory}
Laser spectroscopy experiments allow extracting the nuclear magnetic dipole moment $\mu$ and spectroscopic electric quadrupole moment $Q$ via the hyperfine interaction between the atomic nucleus with spin $I$ and the orbiting electrons with total angular momentum $J$. The hyperfine splitting of an electronic state due to the magnetic dipole (M1) interaction is quantified by the hyperfine $A$-factor as
\begin{equation}\label{eq:DeltaE_M1}
  \Delta E_{\rm M1}(F) = \frac{1}{2}A K,
\end{equation}
where
\begin{equation}\label{eq:K}
  K = F(F+1) - I(I+1) - J(J+1).
\end{equation}
For a point-like nucleus, the hyperfine $A$-factor can be expressed as
\begin{equation}\label{eq:A}
  A = \frac{\mu B_{\rm{el}}}{I J}.
\end{equation}
where $\mu$ is the \textit{point-like} nuclear magnetic dipole moment and $B_{\rm{el}}$ is the magnetic field at the nuclear location due to the electronic motion.

The nuclear magnetic dipole moment can be extracted from measurements of the hyperfine splitting $A$-factor using Eq.~\eqref{eq:A}, if a high-accuracy calculation of the atomic hyperfine magnetic field $B_{\rm{el}}$ is available. In the absence of such calculations and if one assumes that $B_{\rm el}$ is the same for all isotopes of the atomic species (thus neglecting the HFA), $\mu$ can be extracted from $A^{\rm{exp}}$ of the isotope of interest, relative to that of a reference isotope and its independently measured magnetic dipole moment $\mu^{\rm{ref}}$, as
\begin{equation}\label{eq:opticalmu_approx}
  \mu^{\rm{exp}} \approx \frac{I^{\rm{exp}}}{I^{\rm{ref}}}\frac{A^{\rm{exp}}}{A^{\rm{ref}}}\mu^{\rm{ref}}.
\end{equation}

If the hyperfine structure is measured in an atomic level where the electrons penetrate the nuclear volume, then $B_{\rm el}$ is not constant across an isotopic chain. In that case, the expression needs to be corrected by including the differential HFA $^{\rm{ref}}\Delta^{\rm{exp}}$ between the isotope of interest and the chosen reference isotope:

\begin{equation}\label{eq:opticalmu}
  \mu^{\rm{exp}} = \frac{I^{\rm{exp}}}
  {I^{\rm{ref}}}\frac{A^{\rm{exp}}}{A^{\rm{ref}}}\mu^{\rm{ref}}
  \left( 1\,+\,^{\rm{ref}}\Delta^{\rm{exp}} \right).
\end{equation}

Two types of HFA corrections need to be taken into account: the Bohr-Weisskopf (BW) correction $\epsilon_{\rm{BW}}$ and the Breit-Rosenthal (BR) correction. The BW effect is a correction to the magnetic dipole hyperfine splitting due to the finite nuclear magnetization distribution across the nuclear volume~\cite{BohrWeisskopf1950}. That is, the deviation of nuclear magnetization from an ideal point-like magnetic dipole. The absolute value of the BW 
effect for single valence electron states is largest in electronic configurations where the unpaired electron has a large overlap with the nuclear volume~\cite{persson2013}, e.g. in states with $J<1$, such as the $4d^{10}5s\,^2S_{1/2}$ electronic ground state of Ag. However, in some cases when $J>1/2$, the relative BW effect can reach $\sim$10\% due to electron correlation and spin-polarization effects~\cite{maartensson1995, Prosnyak:2020}.

On the other hand, the BR effect in the HFA originates from the extended nuclear charge distribution~\cite{breit1932}. The contribution of the BR effect to the hyperfine splitting is of order 10$\%$ in heavy atoms~\cite{rosenberg1992}.
However, the differential BR correction between two isotopes of the same element is often much smaller than experimental uncertainties, as the charge distribution varies smoothly over an isotopic chain in most cases. According to our calculations, the differential BR correction between $^{97}$Ag and $^{121}$Ag in the electronic ground state is only $0.074\%$, significantly below the experimental uncertainty in all laser-spectroscopic studies of radioactive Ag isotopes so far. Note that the absolute BR effect is taken into account in all calculations due to the use of the extended nuclear charge distribution model.

The experimentally measurable magnetic dipole atomic hyperfine $A$-factor in Eq~\eqref{eq:DeltaE_M1} can be written as the sum of three components~\cite{shabaev1994hyperfine,Skripnikov2020BW}:
\begin{equation}
\label{Aparam}
A = A_0 - A_{\rm{BW}} + A_{\rm{QED}}=(A_{0}+A_{\rm{QED}})(1 - \epsilon),
\end{equation}
where $A_{0}$ is the hyperfine structure constant due to a nucleus with a realistic charge distribution but with an ideal, point-like nuclear magnetic dipole [taking the form of Eq.~\eqref{eq:A}], $A_{\rm{QED}}$ provides a correction to $A_{0}$ due to quantum electrodynamic (QED) effects, $A_{\rm{BW}}$ accounts for the BW effect, and $\epsilon$ is the relative BW correction, 
\begin{equation}
\label{relatBW}
\epsilon=A_{\rm{BW}}/(A_0+A_{\rm{QED}}).
\end{equation}
Note that, in principle, $A^{\rm{QED}}$ also depends on the magnetization distribution. However, we neglect this dependency as $A^{\rm{QED}}$ is already a small correction with respect to other uncertainties for a neutral atom. 

For an electronic state $\left| J,M_J=J \right>$, the point-like magnetic dipole hyperfine factor $A_0$ can be calculated as
\begin{eqnarray}
\label{Apar}
 A_0 = \frac{\mu}{I J}\, 
 \left< J,J \right| \sum_i\, \frac{(-i)}{r_i^2} \sqrt{2} \bm{\alpha} \mathbf{C}^{(0)}_{1,0}(\mathbf{r}_i) \left| J,J \right>,
\end{eqnarray}
where $\mathbf{C}^{(0)}_{1,0}$ is the normalized vector spherical harmonic, $\bm{\alpha}$ represents Dirac matrices, and the matrix element corresponds to the point-like $B_{\rm el}$ from Eq.~\eqref{eq:A}. The finite nuclear magnetization distribution correction $A_{\rm BW}$ is defined as
\begin{eqnarray}
\label{ABWdefinition}
 A_{\rm BW} = \frac{\mu}{I J}\, 
 \left< J,J \right| \sum_i\, \frac{(-i)}{r_i^2} \sqrt{2} \bm{\alpha} \mathbf{C}^{(0)}_{1,0}(\mathbf{r}_i) [1-F(r_i)] \left| J,J \right>,\notag \\
\end{eqnarray} 
where the function $F(r)$ characterizes the radial nuclear magnetization distribution within a finite nucleus.

It was shown in Ref.~\cite{Skripnikov2020BW} that for many-electron heavy atoms and their molecules, even in states with complex electronic structures, the contribution of $A_{\rm BW}$ can be factorized as follows: 
\begin{eqnarray}
 \label{AparBW3}
 A_{\rm BW}
 \approx \frac{\mu}{I J} ({\cal{P}}_s + \beta{\cal{P}}_p)B_s\\
 = \frac{\mu}{I J} \dbtilde{A}_{\rm BW,el} B_s, \nonumber
\end{eqnarray}
where the dimensionless parameter $\dbtilde{A}_{\rm BW,el}$, which depends only on the electronic wavefunction, is introduced as
\begin{equation}
\label{ABwexpr}
\dbtilde{A}_{\rm BW,el}={\cal{P}}_s + \beta{\cal{P}}_p.
\end{equation}
For a given many-electron wave function, ${\cal{P}}_s$ can be calculated as the difference between the mean values of the projectors on the $1s_{1/2}$ hydrogen-like functions with opposite angular momentum projection quantum numbers, truncated at the nuclear charge radius $R_{\rm nuc}$ [see Eqs.~(23), (33), and (34) of Ref.~\cite{Skripnikov2020BW} for details]; we use $R_{\rm nuc}=\sqrt{5/3 \langle r^2\rangle}$, where $\langle r^2\rangle$ is the mean-square nuclear charge radius. ${\cal{P}}_p$ is a similar quantity corresponding to the $2p_{1/2}$ H-like function. The values of ${\cal{P}}_s$ and ${\cal{P}}_p$ can be considered as indicators of the spin-polarization of the $s_{1/2}$ and $p_{1/2}$ electronic shells inside the nucleus.
The parameter $\beta$ represents an electronic factor unique to each element but independent of the atomic state. It can be derived from the ratio of the amplitudes of the large and small components of the $1s_{1/2}$ and $2p_{1/2}$ wave functions within the nucleus~\cite{Shabaev:2006}. Functions with higher total electronic angular momentum ($p_{3/2}$, etc.) are not considered due to their negligible amplitudes inside the heavy nucleus, as described by Eq.~\eqref{ABWdefinition}. In a many-electron state where a single valence electron occupies a $p_{3/2}$ state, the direct BW effect from this electron is minimal. Nonetheless, the effect can manifest via the spin-polarization of ``closed'' $s_{1/2}$ and $p_{1/2}$ electronic shells~\cite{Schwartz:1955,Persson1998}, resulting in non-zero ${\cal{P}}_s$ and ${\cal{P}}_p$ values.

Information about the finite nuclear magnetization distribution can be encapsulated within the nuclear parameter $B_s$ that appears in Eq.~\eqref{AparBW3}, which is defined by the matrix element over the H-like $1s_{1/2}$ wave function $\eta_{1s_{1/2,1/2}}$ with total angular momentum $1/2$ and its projection $1/2$:
\begin{widetext}
\begin{equation}
\label{Bs}
B_s = \int\limits_{|\mathbf{r}|\le R_{\rm nuc}} \eta_{1s_{1/2,1/2}}^{\dagger} \frac{(-i)}{r^2} \sqrt{2} \bm{\alpha} \mathbf{C}^{(0)}_{1,0}(\mathbf{r}) [1-F(r)] \eta_{1s_{1/2,1/2}} d\mathbf{r}.
\end{equation}
\end{widetext}
The electronic factor $\dbtilde{A}_{\rm BW,el}$, as defined in Eq.~\eqref{ABwexpr}, depends solely on the electronic structure, independent of the nuclear magnetization model. This independence has been validated numerically through various nuclear magnetization models for different atoms and molecules~\cite{Skripnikov2020BW,Prosnyak:2021}. Therefore, the parameter $B_s$ encompasses all information about the radial nuclear magnetization distribution $F(r)$ of an isotope, enabling the description of its effect on any electronic state of a many-electron heavy atom, molecule, or compound containing the isotope through Eq.~\eqref{AparBW3}.

There is a well-defined physical meaning of $B_s$; it defines the contribution of the finite nuclear magnetization distribution to the magnetic hyperfine $A$-factor of the H-like ion in the $1s$ electronic ground state (with ${\cal{P}}_s=1$ and ${\cal{P}}_p=0$)~\cite{Skripnikov2020BW,Skripnikov:2022}:
\begin{equation}
\label{BsHlike}
B_s= \frac{I}{2 \mu} A_{\rm BW}({\rm 1s, H\text{-}like~ion}).
\end{equation}

It is convenient to rewrite Eq.~\eqref{Aparam} as follows:
\begin{equation}
\label{Aparam2}
  A = \frac{\mu}{I J}(\Tilde{A}_0 - \dbtilde{A}_{\rm BW,el} B_s + \Tilde{A}_{\rm QED}),
\end{equation}
where we have introduced 
\begin{eqnarray}
  \Tilde{A}_0= \frac{A_0 I J}{\mu},
\end{eqnarray}
corresponding to the point-like hyperfine magnetic field $B_{\rm el}$ from Eq.~\eqref{eq:A}, and 
\begin{eqnarray}
  \Tilde{A}_{\rm QED}= \frac{A_{\rm QED} I J}{\mu}.
\end{eqnarray}
In such a parametrization, $\Tilde{A}_0$, $\dbtilde{A}_{\rm BW,el}$, and $\Tilde{A}_{\rm QED}$ are defined by the electronic wave function and do not depend on the magnetic properties of the nucleus. 

{In the most favorable case, $B_s$ can be extracted from the measurement of the hyperfine $A$-factor in an H-like ion~\cite{Skripnikov:2022,Horst2025}. However, one can also extract it from the hyperfine structure data for a neutral atom, molecule, or an ion. Combining Eqs.~\eqref{AparBW3} and \eqref{Aparam2} provides a blueprint for determining $\mu$ and $B_s$ from laser spectroscopy. Each electronic state in a given species has a magnetic dipole hyperfine structure governed by Eq.~\eqref{Aparam2}, where $\mu/I$ and $B_s$ are specific to the nucleus and $\Tilde{A}_0$, $\dbtilde{A}_{\rm BW,el}$, and $\Tilde{A}_{\rm QED}$ are specific to the electronic state. If the hyperfine $A$-factor is measured in at least two electronic states of the same species, a system of two equations and two unknowns can be constructed to simultaneously extract $\mu/I$ and $B_s$, provided that calculations of $\Tilde{A}_0$, $\dbtilde{A}_{\rm BW,el}$, and $\Tilde{A}_{\rm QED}$ in both electronic states have been performed.

In laser spectroscopy, the hyperfine $A$-factors of both the lower and upper electronic states in an optical transition are measured simultaneously. This means that by performing spectroscopy of a single optical transition, $\mu/I$ and $B_s$ can be extracted~\cite{Skripnikov:2024a} by solving a $2\times2$ equation system of the form
\begin{eqnarray}\label{eq:2x2_1}
  A_1 = (\Tilde{A}_0 + \Tilde{A}_{\rm QED})_1 \frac{\mu}{IJ_1} - (\dbtilde{A}_{\rm BW,el})_1 \frac{\mu}{IJ_1} B_s,\\
  A_2 = (\Tilde{A}_0 + \Tilde{A}_{\rm QED})_2 \frac{\mu}{IJ_2} - (\dbtilde{A}_{\rm BW,el})_2 \frac{\mu}{IJ_2} B_s,\label{eq:2x2_2}
\end{eqnarray}
where the numerical subscripts index electronic states, and $A_1$, $A_2$ are the experimentally measured hyperfine $A$-factors in the two states (see also Ref.~\cite{Porsev:2021}). When this approach is applied to $A_1$ and $A_2$ obtained from a single optical transition, it should be ensured that they are fitted to the hyperfine structure without any relative constraints. If a constraint is placed in the course of the analysis, for instance fixing the ratio $A_1/A_2$, then it should be explicitly considered when solving Eqs.~\eqref{eq:2x2_1}-\eqref{eq:2x2_2}. Even in the absence of a constraint, the $A_1$ and $A_2$ are often correlated; their covariance matrix, when known, should thus be considered in the solution of Eqs.~\eqref{eq:2x2_1}-\eqref{eq:2x2_2} to obtain accurate error bars on $\mu$ and $B_s$.

\begin{table*}
\caption{\label{tab:transitions} Calculated excitation energies (in cm$^{-1}$) for the low-lying electronic states of Ag.}
\begin{tabular}{lllllll}
\hline
\hline
& $5p~^2P^{o}_{1/2}$ & $5p~^2P^{o}_{3/2}$ & $6s~^2S_{1/2}$ & $6p~^2P^{o}_{1/2}$ & $6p~^2P^{o}_{3/2}$ & $4d^95s^2~^2D_{5/2}$  \\
\hline
CCSD(T)             & 29704       & 30649       & 42670     & 48440     & 48653    & 29350  \\
CCSDT-3 $-$ CCSD(T) & 59(19)      & 49(21)      & 178(52)   & 165(58)   & 159(60)  & 703(72)  \\
CCSDT $-$ CCSDT-3   & $-$106(26)    & $-$110(28)    & $-$164(57)  & $-$173(61)  & $-$173(61) & $-$281(63)  \\
CCSDT(Q) $-$CCSDT   & $-$75(75)     & $-$76(76)     & $-$107(107) & $-$110(110) & $-$111(111) & $-$354(354)  \\
Basis set corr.     & 8(8)        & 9(9)        & 9(9)      & 10(10)    & 10(10)    & 674(674) \\
Breit               & $-$19(10)     & $-$31(16)     & $-$47(24)   & $-$46(23)   & $-$49(26)   &  182(91) \\
QED                 & $-$61(18)     & $-$60(18)     & $-$56(17)   & $-$64(19)   & $-$63(19)   & 87(26) \\
Total, this work                  & 29510(85) & 30431(87) & 42484(135) & 48222(142) & 48425(144) & 30361(773)  \\
Ref.~\cite{Ohayon2024Ag} (Theory) & 29484(78) & 30418(79) & 42451(105) & 48190(108) & 48398(103) & --           \\
\\
Experiment~\cite{Badr2004,Civis2011,Badr:2006,Pickering2001}       & 29552.05741 & 30472.66516 & 42556.147 & 48297.406 & 48500.8105 & 30242.298349\\
\hline
\hline
\end{tabular}
\end{table*}

\begin{table*}
\caption{Calculated values of the magnetic dipole hyperfine constants $\Tilde{A}_0$ and $\Tilde{A}_{\rm QED}$ (in MHz/$\mu_N$) for the ground state and excited electronic states.}
\begin{tabular}{lllll}
\hline
\hline
                    & $5s\,^2S_{1/2}$   & $5p\,^2P^{o}_{1/2}$ & $5p\,^2P^{o}_{3/2}$ & $4d^95s^2\,^2D_{5/2}$ \\
\hline                  
DHF                 & 2635.7       & 229.6      & 105.9     & 1305.9   \\
CCSD(T)             & 3897.8       & 384.5      & 201.1     & 1378.2 \\
CCSDT-3 $-$ CCSD(T) & 11.2(12)    & 3.3(6)   & 11.6(3) & 4.0(6)    \\
CCSDT $-$ CCSDT-3   & $-$1.3(4)    & 0.1(2)   & $-$0.6(4) & 0.1(3)    \\
CCSDT(Q) $-$ CCSDT  & 1.7(17)     & 1.0(10)   & $-$0.2(2) & 0.6(6)    \\
Basis-set correction & 1.9(19)     & 0.3(3)  & $-$0.2(2) & 1.2(12)    \\
Breit               & $-$1.9(9)    & $-$1.0(5)  & $-$0.2(1) & 8.2(41)    \\
$\Tilde{A}_{\rm QED}$ & $-$11.3(113)  & 0.0(0)   & $-$0.2(2) & 0.7(7) \\
Total ($\Tilde{A}_0 +\Tilde{A}_{\rm QED}$)               & 3898.2(117) & 388.2(14) & 211.3(6) & 1393.0(44) \\
\hline
\hline
\end{tabular}
\label{TResults}
\end{table*}

If a reliable and precise value of $\mu/I$ has already been measured, for instance via NMR, then $B_s$ can also be determined using the $A$-factor of a single electronic state as
\begin{equation}
\label{BsExpression}
  B_s=
  \frac{\Tilde{A}_0 +\Tilde{A}_{\rm QED} - \frac{I J}{\mu}A^{\rm{exp}}}{ \dbtilde{A}_{\rm BW,el}}.
\end{equation}

Note that alternative parameterizations of the BW effect in many-electron atoms have been suggested in Refs.~\cite{konovalova2017calculation,atoms6030039,Ginges:2022}, and their interrelations are discussed in Ref.~\cite{Skripnikov:2024a}. Related studies on the electronic $g$-factor and hyperfine structure in highly charged ions can also be found in Refs.~\cite{Shabaev:2006,Shabaev:01}.
}

For completeness, the relations between $^a\Delta^b$, $\epsilon$, and $B_s$, where $a$ and $b$ denote two different isotopes of the same element, are provided below:
\begin{align}
  \epsilon &= \frac{A_{\rm BW}}{A_0+A_{\rm{QED}}} = \frac{\dbtilde{A}_{\rm BW,el}}{\Tilde{A}_0 + \Tilde{A}_{\rm QED}} B_s,\label{eq:definition_epsilon}\\
  \frac{A^{(a)}}{A^{(b)}} &= \frac{(A_{0}^{(a)}+A_{\rm{QED}}^{(a)})(1 - \epsilon^{(a)})}{(A_{0}^{(b)}+A_{\rm{QED}}^{(b)})(1 - \epsilon^{(b)})} \nonumber\\
  &=\frac{(A_{0}^{(a)}+A_{\rm{QED}}^{(a)})}{(A_{0}^{(b)}+A_{\rm{QED}}^{(b)})}(1 +\, ^{a}\Delta^{b})\\
  ^{a}\Delta^{b}
    &= \frac{1-\epsilon^{(a)}}{1-\epsilon^{(b)}}-1
     \approx \epsilon^{(b)}-\epsilon^{(a)} \nonumber\\
    &=
    \frac{\dbtilde{A}_{\rm BW,el}}
         {\Tilde{A}_0+\Tilde{A}_{\rm QED}}
    \left(B_s^{(b)}-B_s^{(a)}\right).
    \label{eq:definition_Delta}
\end{align}
 The approximation in Eq.~\eqref{eq:definition_Delta} retains only  terms linear in $\epsilon$. In the last equality, the isotope dependence of the electronic constants $\Tilde{A}_0$, $\Tilde{A}_{\rm QED}$, and $\dbtilde{A}_{\rm BW,el}$ is neglected.
The effect of these isotope dependencies is very small and it is therefore practically always neglected in the discussion of $^{a}\Delta^{b}$ in the literature and in this work.

Since $\epsilon$ and $^{a}\Delta^{b}$ involve electronic factors and $B_s$, they are both isotope-dependent and specific to each electronic state of an atom, ion, or molecule. On the contrary, $B_s$ is a nuclear-dependent parameter that is independent of the electronic state through which it is extracted.

\subsection{Calculated hyperfine fields in the Ag atom}
To benchmark our electronic calculations, we first calculate excitation energies (EE) for six levels in the Ag atom and compare those with precise experimental data and recent calculations from Ref.~\cite{Ohayon2024Ag}, shown in Table~\ref{tab:transitions}. Very good convergence with respect to the basis set size is observed in all cases except for the $5s\,^2S_{1/2} \to 4d^95s^2\,^2D_{5/2}$ transition. In the latter case, the basis‑set correction is significant and is dominated by the $k$‑ and higher‑order harmonics. However, these high‑order harmonics contribute negligibly to the magnetic dipole hyperfine constants (see below). The calculated values of EEs are within 1$\sigma$ of the experimental data~\cite{Badr2004,Civis2011,Badr:2006,Pickering2001}. This reflects the high quality of the employed computational and uncertainty estimation schemes, which were also used in the calculation of hyperfine and isotope-shift constants.

The magnetic dipole hyperfine constants $\tilde{A}_0$ and $\tilde{A}_{\rm QED}$ have been calculated for the $5s\,^2S_{1/2}$ atomic ground state, as well as for the $5p\,^2P^{o}_{1/2}$, $5p\,^2P^{o}_{3/2}$ and $4d^95s^2\,^2D_{5/2}$ excited states, see Table~\ref{TResults}. The results demonstrate good convergence with respect to the basis-set extension and the treatment of high-order correlation effects. Interestingly, there is no near-cancellation of the ``CCSDT-3 $-$ CCSD(T)'' and ``CCSDT $-$ CCSDT-3'' correlation contributions to the hyperfine $A$ constants that is observed in the case of EEs. This trend also holds for the calculations of other properties, such as the electric field gradient (EFG) and the isotope-shift factors considered in later subsections.

For the electronic ground state, the dominant uncertainty for the magnetic dipole hyperfine constant arises from the QED effects. As noted earlier, the current model QED approach considers only certain parts of the necessary QED diagrams. However, the present result for the electronic ground-state QED contribution, $-0.3\%$, is in agreement with the value of $-0.3\%$ that can be obtained by formal interpolation (by $Z$) of QED contributions to the hyperfine constants in neutral alkali atoms with $ns$ electronic ground states~\cite{Sapirstein:2003}.

The calculated values of the $\dbtilde{A}_{\rm BW,el}$ constants [Eq.~\eqref{ABwexpr}], which quantify the sensitivity of each atomic level to the BW effect, are:
\begin{eqnarray}
  \label{ABW,el_s}
  \dbtilde{A}_{\rm BW,el}(5s\,^2S_{1/2})&=&+2.56(1) \times 10^{-4} \\
  \label{ABW,el_p05}
  \dbtilde{A}_{\rm BW,el}(5p\,^2P^{o}_{1/2})&=&+1.75(9) \times 10^{-6} \\
  \label{ABW,el_p15}
  \dbtilde{A}_{\rm BW,el}(5p\,^2P^{o}_{3/2})&=&+1.46(6) \times 10^{-6} \\
  \label{ABW,el_d25}
  \dbtilde{A}_{\rm BW,el}(4d^95s^2\,^2D_{5/2})&=&-3.16(25) \times 10^{-6}.
\end{eqnarray}

The electronic BW sensitivity coefficient, expressed as $\dbtilde{A}_{\rm BW,el} / (\Tilde{A}_0 + \Tilde{A}_{\rm QED})$, is respectively 15, 10 and 29 times smaller for the excited states $5p\,^2P^{o}_{1/2}$, $5p\,^2P^{o}_{3/2}$ and $4d^95s^2\,^2D_{5/2}$ compared to that of the $5s\,^2S_{1/2}$ ground state.

\begin{table*}[htbp]
\caption{\label{ThfsCompare} 
Comparison between experimental and calculated values (in MHz) for the magnetic dipole hyperfine $A$-factors for $^{107}$Ag and $^{109}$Ag in different atomic levels. The calculated values are extracted using Eq.~\eqref{Aparam2}, the calculated electronic structure data of Table~\ref{TResults}, the reference magnetic moments for $^{107}$Ag and $^{109}$Ag determined in this work, and the H-like BW parameters $B_s$.}
\begin{ruledtabular}
\begin{tabular}{lrrrrr}
 & \multicolumn{1}{c}{$5s\,^{2}S_{1/2}$} & \multicolumn{1}{c}{$5p\,^{2}P^{o}_{1/2}$} & \multicolumn{1}{c}{$5p\,^{2}P^{o}_{3/2}$} &   \multicolumn{1}{c}{$4d^9 5s^2\,^2 D_{5/2}$} \\
\midrule  
 & \multicolumn{3}{c}{$^{107}$Ag} \\
 \midrule  
$A_0 + A_{\mathrm{QED}}$ & $-$1770.7(53) & $-$176.3(6) & $-$32.0(1) & $-$126.6(4) \\
$-A_{\mathrm{BW}}$ & 58.2 & 0.4 & 0.1 &  $-$0.1 \\
$A_{\mathrm{th}}$ & $-$1712.5(53)* & $-$175.9(6) & $-$31.9(1) & $-$126.7(4) \\
$A_{\mathrm{exp}}$ & $-$1712.512111(18)~\cite{dahmen1967} & $-$175.4(17)~\cite{Carlsson1990}  & $-$31.7(5)~\cite{Carlsson1990} & $-$126.2818(1)~\cite{blachman1966} \\
      &  &  & $-$32.5(5)~\cite{Bucka:71}  \\
      &  &  & $-$31.7(6)~\cite{Ohayon2024Ag}  \\
\midrule        
& \multicolumn{3}{c}{$^{109}$Ag} \\
\midrule  
$A_0 + A_{\mathrm{QED}}$ & $-$2035.6(61) & $-$202.7(7) & $-$36.8(1)  &  $-$145.5(5) \\
$-A_{\mathrm{BW}}$ & 58.7 & 0.4 & 0.1  & $-$0.1\\
$A_{\mathrm{th}}$ & $-$1976.9(61)* & $-$202.3(7) & $-$36.7(1) & $-$145.6(5) \\
$A_{\mathrm{exp}}$ & $-$1976.932075(17)~\cite{dahmen1967} & $-$201.6(26)~\cite{Carlsson1990} & $-$36.7(7)~\cite{Carlsson1990}  & -145.1584(5)~\cite{blachman1966} \\
      &  &  & $-$37.3(8)~\cite{Bucka:71}  \\
      &  &  & $-$36.9(3)~\cite{Ohayon2024Ag}  \\      
\hline      
\end{tabular}
* The central values coincide with the experimental values by construction. The quoted uncertainties are those of   $A_0 + A_{\mathrm{QED}}$; see text.
\end{ruledtabular}
\end{table*}

\subsection{Bohr-Weisskopf effect in stable isotopes}
After confirming the accuracy of the nuclear magnetic dipole moments of $^{107}$Ag and $^{109}$Ag from the shielding-corrected NMR measurements, the nuclear magnetization distribution parameter $B_s$ can be deduced for the stable isotopes using Eqs.~\eqref{Aparam2},~\eqref{BsExpression} and the experimental $A$-factors from Ref.~\cite{dahmen1967}, $A^{\rm{exp}}(^{107}{\rm Ag})=-1712.512111(18)$~MHz, and $A^{\rm{exp}}(^{109}{\rm Ag})=-1976.932075(17)$~MHz:
\begin{eqnarray}
\label{Bs107NMR}
B_s({\rm ^{107}Ag})= 5.00(1)\{46\}\times 10^5~{\rm MHz}/\mu_N, \\    
\label{Bs109NMR}
B_s({\rm ^{109}Ag})= 4.39(1)\{46\}\times 10^5~{\rm MHz}/\mu_N,
\end{eqnarray}
where the error in parentheses arises from the experimental uncertainty and in curly brackets from the theoretical one.

For isotopes where no independent measurement of the magnetic moment is available, which is the case for most unstable Ag isotopes and isomers, we can solve the $2\times2$ equation system of Eqs.~\eqref{eq:2x2_1},\eqref{eq:2x2_2} to extract $\mu$ and $B_s$ independently. To validate this method, we can apply it for the stable isotopes using the high-precision $A$-factors for the $5s\,^{2}S_{1/2}$ atomic ground state~\cite{dahmen1967} and the $A$-factors for $5p\,^{2}P^{o}_{3/2}$ from Ref.~\cite{Carlsson1990}. The $2\times2$ solution yields $\mu({\rm ^{107}Ag})= -0.1128(20)\{4\}\,\mu_N$  and $\mu({\rm ^{109}Ag})= -0.1307(28)\{4\}\,\mu_N$, in good agreement with but less precise than the values deduced from NMR. The nuclear $B_s$ parameters extracted from the same $2\times2$ solution are also consistent with Eqs.~\eqref{Bs107NMR},\eqref{Bs109NMR}:
\begin{eqnarray}
\label{Bs107-eq}
B_s({\rm ^{107}Ag})= 4.0(26)\{7\}\times 10^5~{\rm MHz}/\mu_N, \\    
\label{Bs109-eq}
B_s({\rm ^{109}Ag})= 4.5(31)\{7\}\times 10^5~{\rm MHz}/\mu_N.
\end{eqnarray}
A more precise measurement of the magnetic dipole hyperfine factors in the $5p\,^{2}P^{o}_{3/2}$ state would dramatically reduce the uncertainties of $\mu$ and $B_s$ from the $2\times2$ solution.

Using the precise values for $B_s$ [Eqs.~\eqref{Bs107NMR},\eqref{Bs109NMR}] and the magnetic dipole moments of Eqs.~\eqref{mu107},\eqref{mu109}, we can now calculate the different contributions to the magnetic dipole hyperfine $A$-factor in the excited electronic states for $^{107}$Ag and $^{109}$Ag, shown in Table~\ref{ThfsCompare}. It is found that the theoretical $A$-factors ($A_{\rm th}$) are in agreement within 2$\sigma$ with experiment for both isotopes and all excited electronic states (for the ground state, the agreement is exact by construction). The theoretical values for the hyperfine $A$-factors of the $5p\,^2P^{o}_{1/2}$ and $5p\,^2P^{o}_{3/2}$ electronic states have smaller uncertainty than the currently available experimental measurements.

\begin{table*}[]
\caption{Summary of the presently extracted nuclear magnetic dipole moments ($\mu$) and absolute hydrogen-like Bohr-Weisskopf nuclear magnetization distribution parameters ($B_s$) for all Ag isotopes with measured hyperfine $A$-factors in literature. The listed moments for $^{97-100}$Ag are calculated using the Moskowitz-Lombardi rule~\cite{moskowitz1973}, for $^{107-109}$Ag they are the re-evaluated NMR values, and for $^{113m-121}$Ag they are calculated using Eqs.~\eqref{eq:2x2_1}-\eqref{eq:2x2_2}. For $^{101-105m}$Ag, the moments are the original values reported in Ref.~\cite{dinger1989} scaled by the corrected reference moment $\mu(^{109}{\rm Ag})$ from this work, but uncorrected for the HFA that is considered to have an effect significantly below the error bar. Uncertainties originating from the experimental and theoretical contributions are given in parentheses and curly brackets, respectively. The spins and parities are taken from the ENSDF database~\cite{ENSDF} unless specified otherwise. For $^{98}$Ag, two different potential spin assignments are considered.} \label{tab:new_dipole_moments}
\begin{tabular}{cccccl}
\toprule
\toprule
Isotope   & $I^\pi$ &  \multicolumn{2}{c}{$A(^2S_{1/2})$ (MHz)}  & \hspace{0.4cm}$\mu$~($\mu_N$)    & $B_s$~($10^5$\,MHz/$\mu_N$)    \\\hline
97  & (9/2$^+$) &  \multicolumn{2}{c}{+10600(200)~\cite{ferrer2014}} &+6.12(12)                          & \\
98  & (6$^+$)   &  \multicolumn{2}{c}{+6020(90)~\cite{ferrer2014}} &+4.64(7)                           &  \\
    & (5$^+$)   &  \multicolumn{2}{c}{+7120(110)~\cite{ferrer2014}} &+4.57(7)                           &  \\
99  & (9/2)$^+$ &  \multicolumn{2}{c}{+10050(50)~\cite{ferrer2014}} &+5.80(3)                           &  \\
100 & (5)$^+$   & \multicolumn{2}{c}{+6810(40)~\cite{ferrer2014}} &+4.37(3)                           &  \\
\toprule
Isotope   & $I^\pi$ &  \multicolumn{2}{c}{$A(^2D_{5/2})$ (MHz)}  & \hspace{0.4cm}$\mu$~($\mu_N$)    & $B_s$~($10^5$\,MHz/$\mu_N$)    \\\hline
101 & 9/2$^+$   & \multicolumn{2}{c}{+694.4(14)~\cite{dinger1989}}  &+5.621(11)                         &   \\
103 & 7/2$^+$   &  \multicolumn{2}{c}{+703.2(14)~\cite{dinger1989}} &+4.427(9)                          &  \\
104 & 5$^+$     &  \multicolumn{2}{c}{+435.3(6)~\cite{dinger1989}} &+3.915(5)                          &  \\
105m& 7/2$^+$   &  \multicolumn{2}{c}{+700.4(24)~\cite{dinger1989}} &+4.409(15)                         &  \\
\toprule
Isotope   & $I^\pi$ &  \multicolumn{2}{c}{$A(^2S_{1/2})$ (MHz)}  & \hspace{0.4cm}$\mu$~($\mu_N$)    & $B_s$~($10^5$\,MHz/$\mu_N$)    \\\hline
107 & 1/2$^-$   &  \multicolumn{2}{c}{$-$1712.512111(18)~\cite{dahmen1967}} &$-$0.113558(9)                       & +5.00(1)\{46\} \\
109 & 1/2$^-$   &  \multicolumn{2}{c}{$-$1976.932075(17)~\cite{dahmen1967}} &$-$0.130551(10)                      & +4.39(1)\{46\} \\
\toprule
Isotope   & $I^\pi$ &  \hspace{0.4cm}$A(^2S_{1/2})$ (MHz) & \hspace{0.4cm}$A(^2P_{3/2})$ (MHz)  & \hspace{0.4cm}$\mu$~($\mu_N$)    & $B_s$~($10^5$\,MHz/$\mu_N$)    \\\hline
113m& 7/2$^+$   & +9609(5)~\cite{degroote2024AgPLB} & +173.7(13)~\cite{degroote2024AgPLB,vandenborne2025} &+4.316(36)\{14\}                     & +0.1(13)\{7\} \\
114 & 1$^+$     & +19215(22)~\cite{vandenborne2025} & +352(4)~\cite{vandenborne2025} & +2.503(32)\{8\}                     & +2.3(19)\{7\} \\
115m& 7/2$^+$   & +9556(2)~\cite{degroote2024AgPLB} & +173.5(7)~\cite{degroote2024AgPLB,vandenborne2025} & +4.313(19)\{14\}                    & +0.8(7)\{7\} \\
116m1& $^\bullet$4$^+$  & +5329(5)~\cite{vandenborne2025} & +96.6(10)~\cite{vandenborne2025}  & +2.744(32)\{9\}           & +0.5(18)\{7\} \\
116m2& $^\bullet$7$^-$  & +3304.0(5)~\cite{vandenborne2025} & +60.10(12)~\cite{vandenborne2025}  & +2.989(7)\{10\}            & +1.1(3)\{7\} \\
117m& $^\bullet$7/2$^+$ & +9486(2)~\cite{degroote2024AgPLB} & +172.5(4)~\cite{degroote2024AgPLB,vandenborne2025}  &+4.289(11)\{14\}           & +1.1(4)\{7\} \\
118 & $^\bullet$4$^+$   & +5911(2)~\cite{vandenborne2025} & +106.0(12)~\cite{vandenborne2025}  & +3.007(38)\{10\}           & $-$1.3(19)\{7\}  \\
118m2& $^\bullet$7$^-$  & +3812.9(10)~\cite{vandenborne2025} & +68.8(4)~\cite{vandenborne2025} & +3.418(22)\{11\}          & $-$0.2(10)\{7\} \\
119m& $^\bullet$7/2$^+$ & +9581(2)~\cite{degroote2024AgPLB} & +177.2(5)~\cite{degroote2024AgPLB,vandenborne2025} &+4.415(14)\{14\}$^\star$   & +3.9(5)\{7\}$^\star$ \\
120 & $^\bullet$4$^+$   & +6045(4)~\cite{vandenborne2025} & +107.9(16)~\cite{vandenborne2025} & +3.059(51)\{10\}           & $-$2.1(26)\{7\} \\
120m2&$^\bullet$7$^-$   & +3651(3)~\cite{vandenborne2025} & +65.5(9)~\cite{vandenborne2025} & +3.252(50)\{10\}          & $-$1.3(24)\{7\} \\
121 & $^\bullet$7/2$^+$ & +9610(3)~\cite{degroote2024AgPLB} & +172.5(16)~\cite{degroote2024AgPLB,vandenborne2025} & +4.283(44)\{14\}          & $-$1.1(16)\{7\} \\
\hline\hline
\end{tabular}
\begin{tablenotes}
\item ($\bullet$) Spin assignment from Refs.~\cite{degroote2024AgPLB,vandenborne2025}.
\item $(\star)$  Possible outlier (see text for details).
\end{tablenotes}
\end{table*}

\begin{figure*}[htb]
    \centering
    \includegraphics[width=0.7\linewidth]{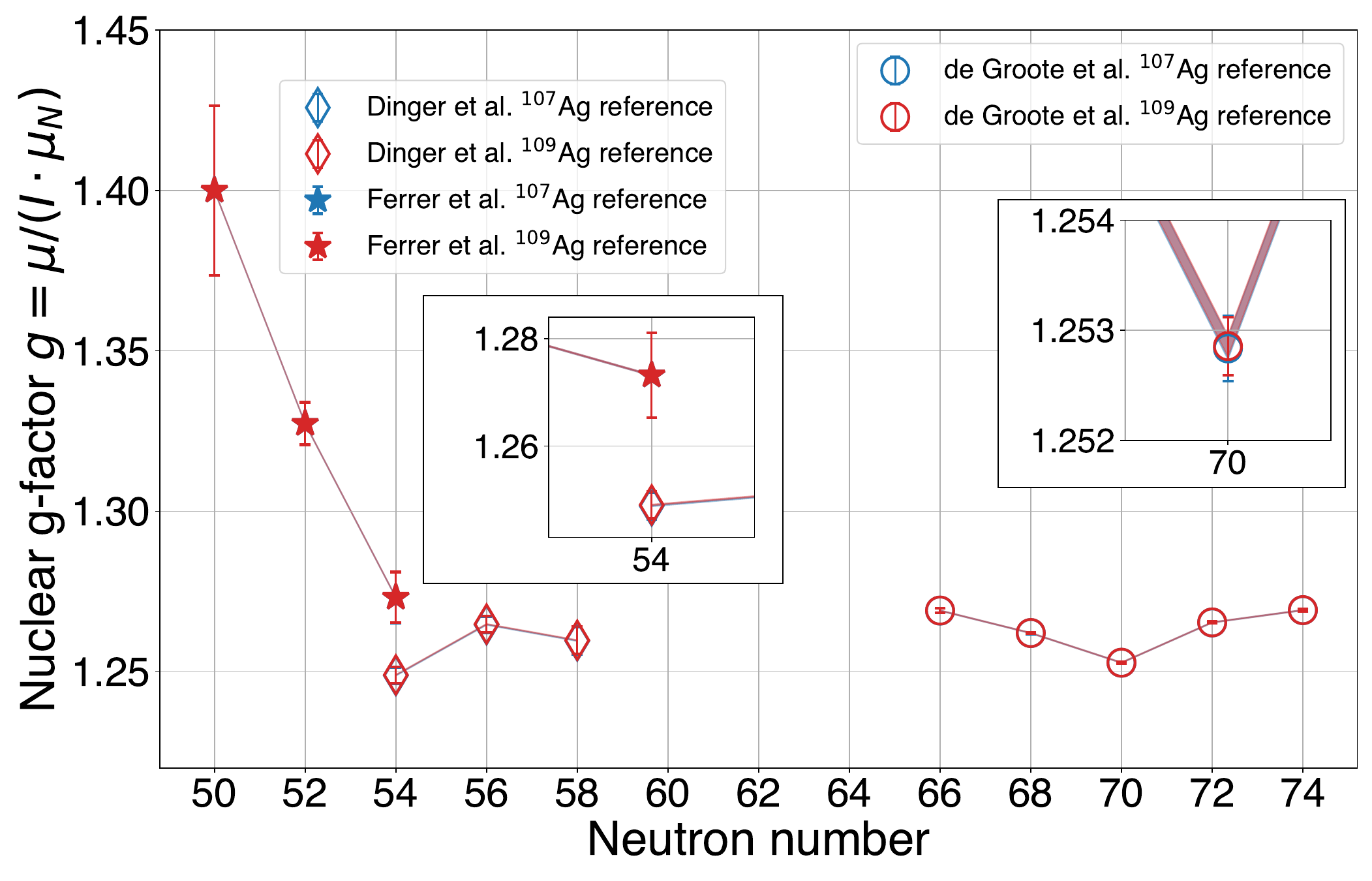}
    \caption{Overview of nuclear $g$-factors of odd-even Ag isotopes with spin $I = 7/2^+$ ($I=9/2^+$ for $^{97,99,101}$Ag) from laser spectroscopy. Error bars show the experimental statistical uncertainty. The uncertainties originating from the reference moments are given by the colored band, visible as a line on the figure. Hyperfine parameters reported by Ferrer \textit{et al.}~\cite{ferrer2014}, Dinger \textit{et al.}~\cite{dinger1989} and de Groote \textit{et al.}~\cite{degroote2024AgPLB} were used to extract $g$-factors relative to two reference isotopes using Eq.~\eqref{eq:opticalmu}, including the HFA between $^{107,109}$Ag. The reference magnetic moments were taken from this work in Table~\ref{tab:NMR_comparison}.}
    \label{fig:new_Lit_overview_g-factor}
\end{figure*}

The experimental differential HFA that can be deduced with Eq.~(4) from Ref.~\cite{persson2013} using the precisely measured $A$-factors in Table~\ref{tab:new_dipole_moments} and the reference moments [Eq.~\eqref{mu107}-\eqref{mu109}], giving $^{107}\Delta^{109} = -0.41277(7)\%$ for $5s\,^2S_{1/2}$ and $^{107}\Delta^{109} = +0.0139(4)\%$ for $4d^9 5s^2\,^2 D_{5/2}$.
In the present approach, the differential HFA for the $5s\,^2S_{1/2}$ state does not depend on the calculated electronic constants $\Tilde{A}_0 + \Tilde{A}_{\rm QED}$  and $\dbtilde{A}_{\rm BW,el}$, whereas that for the $4d^95s^2\,^2D_{5/2}$ state does. Using the results of Eqs.~\eqref{Bs107NMR} and~\eqref{Bs109NMR}, together with those of Eqs.~\eqref{ABW,el_s} and~\eqref{ABW,el_d25}, we obtain a theoretical value of $^{107}\Delta^{109}=+0.0138(11)\%$ for the $4d^95s^2\,^2D_{5/2}$ state, in good agreement with the experimental value. This agreement tests the accuracy of the calculated $\dbtilde{A}_{\rm BW,el}/\Tilde{A}_0$ ratio for the $4d^95s^2\,^2D_{5/2}$ electronic state.
Taking the experimentally determined HFA for the $5s\,^2S_{1/2}$ level and the re-evaluated moments of $^{107}$Ag and $^{109}$Ag, we find that the inconsistency of Fig.~\ref{fig:lit_g-factor} between reference isotopes is resolved, as shown in Fig.~\ref{fig:new_Lit_overview_g-factor}. We observe no statistically significant deviation in the magnetic dipole moments measured by Ferrer \textit{et al.} and de Groote \textit{et al.} for the two reference isotopes. Note that an additional error is introduced when $^{107}$Ag is used as a reference due to the uncertainty in the differential HFA ($^{107}\Delta ^{109}$). However, there remains a disagreement on the magnetic dipole moment of $^{101}$Ag ($N=54$), which was measured by Ferrer \textit{et al.} in the $5s\,^2S_{1/2}$ atomic state and by Dinger \textit{et al.} in the $4d^95s^2\,^2D_{5/2}$ state. Likely, the disagreement is due to the effect of the HFA between $^{101}$Ag and the reference isotopes $^{107}$Ag and $^{109}$Ag, which has a significant contribution in $5s\,^2S_{1/2}$ but a negligible one in $4d^95s^2\,^2D_{5/2}$. For example, an HFA of $^{107}\Delta^{101}\approx$~$^{109}\Delta^{101}\approx -1.8\%$ would bring the $g$-factor of $^{101}$Ag extracted from data by Ferrer \textit{et al.} and Dinger \textit{et al.} into agreement. However, due to the low precision in the data by Ferrer \textit{et al.}, it is not possible to determine the values for $^{107}\Delta^{101}$ and $^{109}\Delta^{101}$ with an uncertainty of less than 100$\%$. As a result, there is a need for high-precision laser spectroscopy of $^{97-105}$Ag in the atomic $5s\,^2S_{1/2}$ and $5p\,^2P_{1/2}$ or $5p\,^2P_{3/2}$ levels to determine the HFA for these isotopes.

\subsection{Nuclear magnetic dipole moments of radioactive isotopes}
As noted above, all laser-spectroscopic studies so far have reported nuclear magnetic dipole moments of radioactive Ag isotopes by scaling the measured $A$-factors of the isotopes under study with the magnetic dipole moment and $A$-factor of a reference isotope -- often $^{109}$Ag.

We have shown, however, that a discrepancy remains between the magnetic dipole moments extracted from hyperfine structure data measured in the $4d^9\,5s^2\,^2 D_{5/2}$ and $5s\,^{2}S_{1/2}$ levels (see Fig.~\ref{fig:new_Lit_overview_g-factor}). As explained above, this is most likely due to the large BW effect in the $5s\,^{2}S_{1/2}$ level, which induces a systematic uncertainty on the magnetic moments extracted from $A(^2S_{1/2})$ data. As the BW effect in the excited electronic states of Ag considered here is at least an order of magnitude smaller than in the $5s\,^{2}S_{1/2}$ state, reliable magnetic moments can be deduced from the hyperfine structure in those states using the measured $A$-factors and the calculated values of $\Tilde{A}_0 +\Tilde{A}_{\rm QED}$ for $5p\,^2P^{o}_{1/2}$, $5p\,^2P^{o}_{3/2}$ or $4d^95s^2\,^2D_{5/2}$ listed in Table~\ref{TResults}, neglecting the $\dbtilde{A}_{\rm BW,el}B_s$ term.

As a consistency test for this approach, the magnetic dipole moments of the stable $^{107,109}$Ag can be compared when extracted via the hyperfine $A$-factors in $5p\,^2P^{o}_{1/2}$~\cite{Carlsson1990}, $5p\,^2P^{o}_{3/2}$~\cite{Carlsson1990}, and $4d^95s^2\,^2D_{5/2}$~\cite{blachman1966}, using Eq.~\eqref{Aparam2}:
\begin{eqnarray}
\mu({\rm ^{107}Ag},~^2P^o_{1/2})&=& -0.1130(11)\{4\} \mu_N,\\
\mu({\rm ^{107}Ag},~^2P^o_{3/2})&=&-0.1125(18)\{3\} \mu_N, \\
\mu({\rm ^{107}Ag},~^2D_{5/2})&=&-0.1133\{4\} \mu_N,
\end{eqnarray}
\begin{eqnarray}
\mu({\rm ^{109}Ag},~^2P^o_{1/2})&=& -0.1298(17)\{5\} \mu_N, \\
\mu({\rm ^{109}Ag},~^2P^o_{3/2})&=&-0.1303(25)\{4\} \mu_N, \\
\mu({\rm ^{109}Ag},~^2D_{5/2})&=&-0.1303\{4\} \mu_N.
\end{eqnarray}
The uncertainties in parentheses originate from the statistical error on the measured $A$-factors in the excited electronic states. For the $^2D_{5/2}$ data, the statistical uncertainty is three orders of magnitude smaller than the theoretical one (in curly brackets) due to the sub-kHz precision of the experimental $A$-factors~\cite{blachman1966}, and thus they are not shown. All values are consistent with each other, as well as with our re-evaluated reference moments derived from NMR data. This demonstrates that magnetic moments derived from these electronic levels, which are less sensitive to a systematic error due to the unknown BW effect, are reliable~\footnote{We note that the BW contribution induces a systematic uncertainty to the value of $\mu$ extracted with this method that is significantly smaller than the current experimental uncertainty in the published $A$-factors. It is only for the stable isotopes that the BW effect in the $4d^95s^2\,^2D_{5/2}$ state exceeds the experimental uncertainty, although it remains below the theoretical uncertainty in this case.}.

Our extracted nuclear magnetic dipole moments $\mu$ and nuclear magnetization distribution parameters $B_s$ of the radioactive Ag isotopes are listed in Table~\ref{tab:new_dipole_moments}. For $^{97-100}$Ag, the low precision in experimental data leads to $B_s$ values with uncertainty in the order of 100$\%$, so they are not tabulated. The magnetic dipole moments of these isotopes are calculated with Eq.~\eqref{eq:opticalmu} using the re-evaluated moment of $^{109}$Ag [Eq.~\eqref{mu109}] and the HFA approximated by the Moskowitz-Lombardi rule as in Ref.~\cite{ferrer2014}. For $^{101,103,104,105m}$Ag, the moments are the originally reported values from Ref.~\cite{dinger1989} scaled by the corrected reference moment $\mu(^{109}{\rm Ag})$ from this work, but uncorrected for the HFA that is considered to be negligible at the current level of precision. For the neutron-rich isotopes ($A>109$), $\mu$ and $B_s$ values are extracted using the measured $A$-factors in the $5p\,^2P^{o}_{3/2}$ and $5s\,^2S_{1/2}$ states from the solution of the $2\times2$ equation system of Eqs.~\eqref{eq:2x2_1},\eqref{eq:2x2_2}. The extracted $B_s$ parameters are discussed in the next section.

\begin{figure*}[htb]
    \centering
    \includegraphics[width=0.7\linewidth]{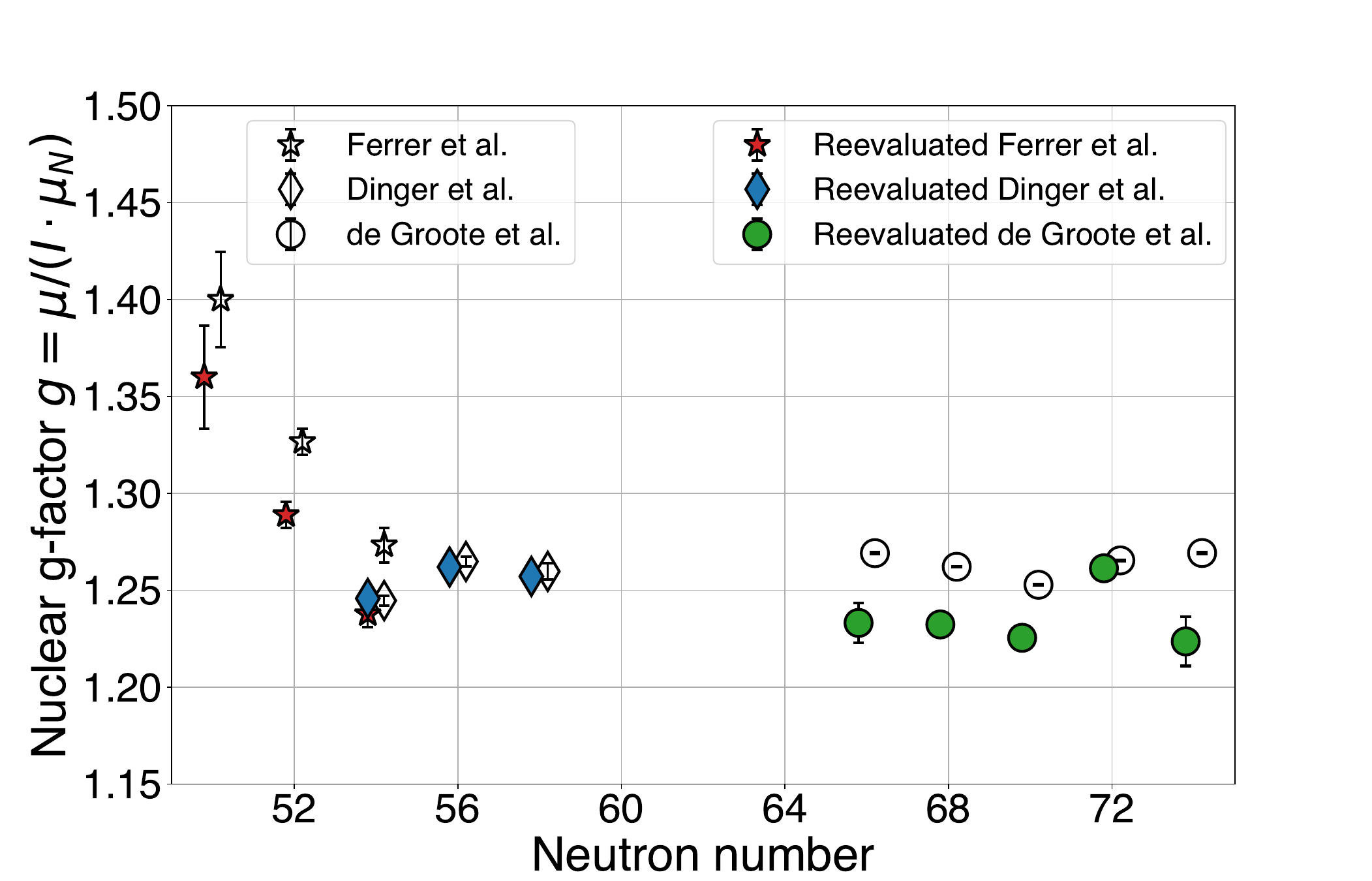}
    \caption{Overview of the nuclear $g$-factors for odd-even Ag isotopes with spin $I = 7/2^+$ ($I=9/2^+$ for $^{97,99,101}$Ag) from laser spectroscopy. The empty markers correspond to the originally published $g$-factors as reported in the literature. The solid markers correspond to the $g$-factors determined as $g=\mu/(I\mu_N)$ from Table~\ref{tab:new_dipole_moments}. A horizontal offset between different markers corresponding to the same isotope is added for visual clarity.}
    \label{fig:recommended_Lit_overview_g-factor}
\end{figure*}

Figure~\ref{fig:recommended_Lit_overview_g-factor} shows a comparison between the originally reported nuclear $g$-factors for the spin-7/2(9/2) states in the Ag isotopic chain and the $g$-factors extracted in this work. When explicitly taking into account the BW effect in the $^2 S_{1/2}$ state, the corrected Ferrer \textit{et al.} and Dinger \textit{et al.} $g$-factors for $N=54$ are in agreement. On the other hand, the nuclear $g$-factors extracted from the present analysis show an unexpected jump at $^{119m}$Ag, which is also reflected in the resulting $B_s$ values (see next section). It is unclear whether this abrupt jump in the nuclear $g$-factor for the spin-7/2 state at $N=72$ corresponds to a real nuclear-structure effect, or whether it originates from an analysis artifact on the reported $A$-factors. We do not attempt to provide a physics interpretation of this observation in this work, and thus we do not claim that the presently extracted magnetic dipole moment should be seen as more accurate or reliable than the value reported in Ref.~\cite{degroote2024AgPLB}. We call for a re-measurement of this isotope to clarify the origin of this observation.

Further laser-spectroscopic experiments are also needed to expand the list of Ag isotopes for which the hyperfine splitting in $5p\,^2P^{o}_{3/2}$, $5p\,^2P^{o}_{1/2}$, or $4d^95s^2\,^2D_{5/2}$ is measured with high precision, especially for the low-spin nuclear states, for which only the ground states of the stable $^{107,109}$Ag have been reported so far. It should be noted that the $5p\,^2P^{o}_{1/2}$ state is preferable for extracting the magnetic dipole moment, as the $A$-factor is approximately a factor of 5.5 larger than in $5p\,^2P^{o}_{3/2}$~\cite{Carlsson1990}. Therefore, the hyperfine splitting will be easier to resolve and thus have a smaller relative uncertainty, leading to a more precise magnetic dipole moment.

One can find that in the $2\times2$ equation method, the uncertainty of the present values of magnetic dipole moments is larger than in previous studies (see Table~\ref{tab:new_dipole_moments}). Note, however, that this is due to including the excited electronic state, which has a lower relative precision, to extract the magnetic moment. In the present study, we opted to use a combined solution [Eqs.~\eqref{eq:2x2_1},\eqref{eq:2x2_2}] that includes hyperfine factors for $5p\,^2P^{o}_{3/2}$, as the HFA is practically negligible compared to experimental uncertainties for this state. 

\subsection{Bohr-Weisskopf effect in radioactive isotopes}
For many isotopes in Table~\ref{tab:new_dipole_moments}, including all neutron-deficient isotopes, the value of $B_s$ is consistent with zero due to the large uncertainties in the hyperfine $A$-factors, measured with broadband laser spectroscopy.

In the absence of reported covariance matrices between the fitted $A$-factors in the laser spectroscopy literature, the extracted $\mu$ and $B_s$ in Table~\ref{tab:new_dipole_moments} do not take into account the covariance between the two $A$-factors used in the $2\times2$ system of equations. The $A$-factors of the lower and upper electronic states were thus considered to be independent. The calculated electronic-structure constants in the upper and lower states are also assumed to be independent, giving rise to the systematic errors in Table~\ref{tab:new_dipole_moments}.

By construction, the values of $B_s$ in Eqs.~\eqref{Bs107NMR},\eqref{Bs109NMR} and Table~\ref{tab:new_dipole_moments} are consistent with the HFA values, such as $^{107}\Delta^{109}=-0.413(11)$\%~\cite{burges1973} for the electronic ground state $5s\,^2S_{1/2}$ of Ag. However, in this work, the values of the BW corrections $\epsilon$ can be directly determined for each isotope individually, rather than only the relative values between isotopes. This approach in principle allows probing nuclear structure calculations of the magnetization distribution function $F(r)$ for a given isotope through Eq.~\eqref{Bs}.

\begin{figure}
    \centering
    \includegraphics[width=0.5\textwidth]{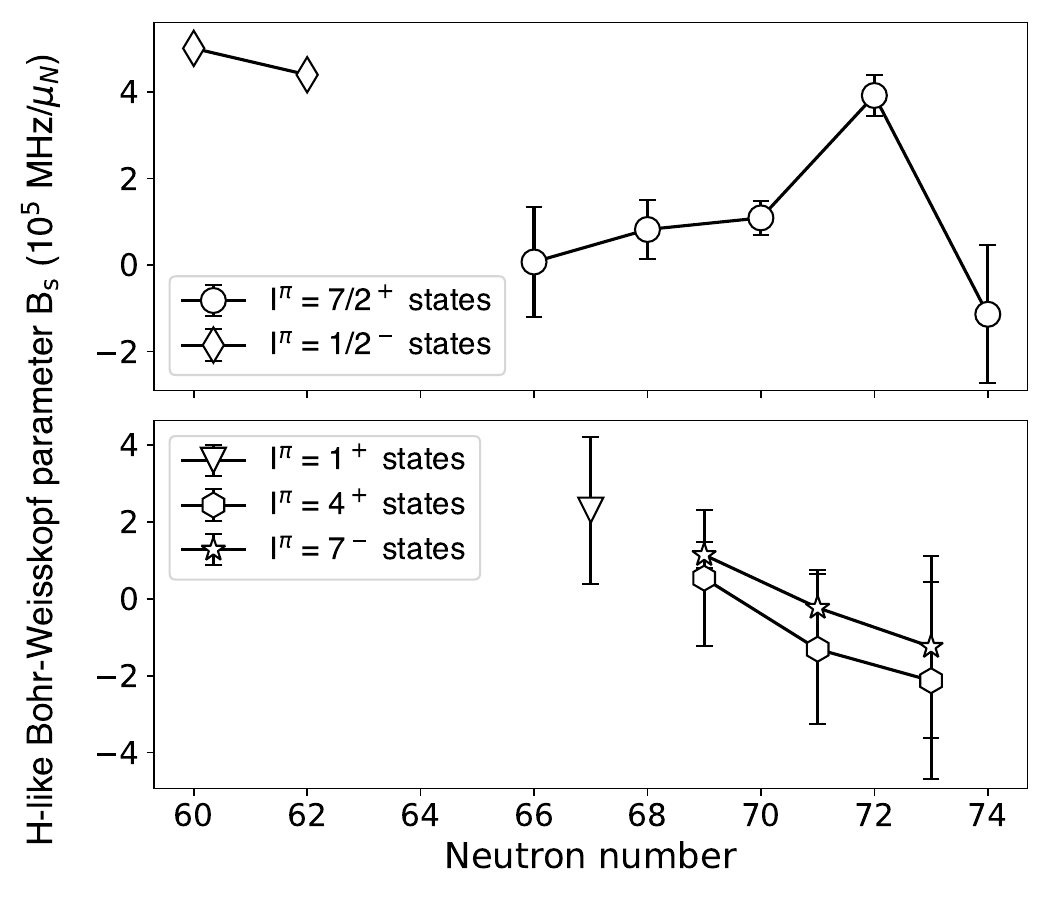}
    \caption{Evolution of the hydrogen-like Bohr-Weisskopf nuclear parameter $B_s$ for the $I^\pi=1/2^-,\,7/2^+$ and $I^\pi=1^+,\,4^+,\,7^-$ nuclear states in Ag isotopes, as extracted in this work and listed in Table~\ref{tab:new_dipole_moments}. The theoretical uncertainties listed in Table~\ref{tab:new_dipole_moments} are not shown.}
    \label{fig:Bs_trend}
\end{figure}

Following the definition in Eq.~\eqref{eq:definition_epsilon}, the BW parameters $\epsilon$ for $^{107,109}$Ag are: $\epsilon({\rm ^{107}Ag}, 5s\,^2S_{1/2})=+3.284(7)\{301\}$\% and $\epsilon({\rm ^{109}Ag}, 5s\,^2S_{1/2})=+2.885(7)\{301\}$\%. The latter value can be compared with the estimates of $\epsilon({\rm ^{109}Ag})$ performed within the single-valence-nucleon nuclear models using the uniform spherical nuclear magnetization distribution, which yields $\epsilon_{\rm spherical}(^{109}{\rm Ag})=+3.3\%$, and the trapezoidal nuclear magnetization distribution, which yields $\epsilon_{\rm trapezoidal}(^{109}{\rm Ag})=+3.8\%$~\cite{schmelling1967}. Note that the uncertainty of simplistic models, such as the uniform spherical distribution model, can be 30\% or higher~\cite{schmelling1967,shabaev1994hyperfine}. 

Figure~\ref{fig:Bs_trend} shows the values for the $B_s$ parameter determined for Ag isotopes in this work, as listed in Table~\ref{tab:new_dipole_moments}. While the uncertainties for most data points are too large to draw conclusions relevant to nuclear structure, the $I^\pi=1/2^-$ stable isotopes exhibit substantially larger extracted $B_s$ values than most of the neutron-rich high-spin states; additional data are required to determine whether this reflects a systematic dependence on nuclear configuration.

\section{Nuclear electric quadrupole moments}

We now turn to the electric quadrupole moment $Q$. Due to the absence of a stable isotope with $I>1/2$, the only way to extract $Q$ is by combining measurements on radioactive isotopes with calculations of electric field gradients at the site of the nucleus. We perform these calculations for the atomic system. 

\subsection{Electric quadrupole hyperfine interaction theory}
The hyperfine splitting due to the electric quadrupole (E2) interaction is quantified by the hyperfine $B$-factor as:
\begin{equation}\label{eq:DeltaE_E2}
  \Delta E_{\rm E2}(F) = B \frac{  3K(K+1) - 4I(I+1)J(J+1) }{8I(2I-1)J(2J-1)},
\end{equation}
where
\begin{equation}\label{eq:B}
  B = eQV_{zz}
\end{equation}
with $V_{zz}$ being the EFG formed by the electrons at the location of the nucleus, and $e$ is the elementary charge. The EFG can be calculated as:
 \begin{equation}   
   V_{zz}=2\langle J,J | \sum_i -\frac{1}{r_i^3} C_{2,0}(\mathbf{r}_i) | J,J\rangle,
\end{equation}
where $C_{2,0}$ is the zeroth component of normalized spherical harmonic. 

In the absence of an accurate calculation of the EFG, the spectroscopic electric quadrupole moments $Q^{\rm{exp}}$ can also be determined from optical measurements in a manner similar to the extraction of $\mu^{\rm{exp}}$. The measured quadrupole hyperfine factor $B^{\rm{exp}}$ can be used in combination with the hyperfine factor $B^{\rm{ref}}$ and electric quadrupole moment $Q^{\rm{ref}}$ for a reference isotope, with $Q^{\rm{ref}}$ often extracted from non-optical measurements. The electric quadrupole moment of any isotope $Q^{\rm{exp}}$ can then be extracted as:
\begin{equation}\label{eq:opticalQ}
    Q^{\rm{exp}} = \frac{B^{\rm{exp}}}{B^{\rm{ref}}} Q^{\rm{ref}}
\end{equation}
It should be noted, however, that quadrupole moments from non-optical measurements are often determined from the interaction of the nucleus with an EFG in a crystal, which is typically even less accurately determined than EFG calculations in atoms.

Our accurate and precise atomic coupled cluster calculations in this work aim to provide, for the first time, accurate quadrupole moments of a wide range of Ag isotopes using the high-precision measurements obtained with laser spectroscopy in the $4d^{10}5p\,^2P_{3/2}$~\cite{degroote2024AgPLB,vandenborne2025} and $4d^{9}5s^2\,^2D_{5/2}$~\cite{dinger1989} atomic states.

\begin{table}
\caption{Electric quadrupole hyperfine factors ($B$) and literature quadrupole moments ($Q_{\rm lit}$) from measurements in the atomic $4d^{10}5p\,^2P^{o}_{3/2}$ (Refs.~\cite{vandenborne2025,degroote2024AgPLB,fischer1975}) and $4d^9 5s^2\,^2 D_{5/2}$ (Ref.~\cite{dinger1989}) states, along with the quadrupole moments ($Q_{\rm new}$) determined in this work. The extraction was carried out following Eq.~\eqref{eq:B} using the tabulated values of $B$ along with the calculated electric field gradient for each atomic state from Table~\ref{TResultsEFG}. Uncertainties from experimental and theoretical contributions are given in parentheses and curly brackets, respectively. Quadrupole moments of $^{107m,109m}$Ag from level-mixing resonance on oriented nuclei are given for completeness but are not re-evaluated. The electric quadrupole moment of $^{110m}$Ag, which is often the reference moment of choice to extract the quadrupole moments of other isotopes, is given in bold.}\label{tab:Q_moments}
\begin{tabular}{lccc}
\hline\hline
Isotope   & \hspace{0.3cm}$B$~(MHz)\hspace{0.3cm} & \hspace{0.4cm}$Q_{\rm{lit}}$~($e$b)\hspace{0.4cm}  & \hspace{0.4cm}$Q_{\rm{new}}$~($e$b)\hspace{0.4cm} \\\midrule
\multicolumn{4}{c}{Laser spectroscopy: $4d^95s^2\,^2D_{5/2}$~\cite{dinger1989}} \\
101 & $-$448(30) & +0.33(5) & +0.372(25)\{0\} \\
103 & $-$1079(18) & +0.80(9) & +0.897(15)\{1\} \\
104 & $-$1364(15) & +1.01(11) & +1.134(13)\{1\} \\
105m & $-$1099(41) & +0.81(11) & +0.913(34)\{1\} \\
106m & $-$1431(94) & +1.06(16) & +1.189(78)\{1\} \\\midrule
\multicolumn{4}{c}{Level-mixing resonance on oriented nuclei} \\
107m & & +0.98(11)~\cite{berkes1986} & \\
109m & & +1.02(12)~\cite{berkes1986} & \\\midrule
\multicolumn{4}{c}{Laser spectroscopy: $4d^{10}5p\,^2P^{o}_{3/2}$} \\
108m & +393(15)~\cite{fischer1975}  & +1.32(7)~\cite{berkes1984} & +1.389(53)\{5\} \\
110m & +425(18)~\cite{fischer1975}  & +1.44(10)~\cite{berkes1984}       & \textbf{+1.502(64)\{5\}} \\
113m & +305(12)~\cite{degroote2024AgPLB}& +1.03(9)~\cite{degroote2024AgPLB} & +1.078(42)\{4\} \\
114 & +59(3)~\cite{vandenborne2025} & +0.201(11)\{16\}~\cite{vandenborne2025}& +0.209(11)\{1\} \\
115m  & +309(6)~\cite{degroote2024AgPLB} & +1.04(8)~\cite{degroote2024AgPLB} & +1.092(21)\{4\} \\
116m1 & +234(38)~\cite{vandenborne2025}& +0.79(13)\{6\}~\cite{vandenborne2025} & +0.827(134)\{3\} \\
116m2 & +293(7)~\cite{vandenborne2025} & +0.99(3)\{8\}~\cite{vandenborne2025}  & +1.036(25)\{3\} \\
117m  & +309(5)~\cite{degroote2024AgPLB} & +1.05(8)~\cite{degroote2024AgPLB} & +1.092(18)\{4\} \\
118   & +297(13)~\cite{vandenborne2025} & +1.00(4)\{8\}~\cite{vandenborne2025} & +1.050(46)\{4\} \\
118m2 & +408(6)~\cite{vandenborne2025} & +1.38(2)\{11\}~\cite{vandenborne2025} & +1.442(21)\{5\} \\
119m  & +276(7)~\cite{degroote2024AgPLB} & +0.93(8)~\cite{degroote2024AgPLB} & +0.976(25)\{3\} \\
120   & +299(17)~\cite{vandenborne2025} & +1.01(6)\{8\}~\cite{vandenborne2025} & +1.057(60)\{4\} \\
120m2 & +371(7)~\cite{vandenborne2025} & +1.25(3)\{10\}~\cite{vandenborne2025} & +1.311(25)\{4\} \\
121   & +249(10)~\cite{degroote2024AgPLB}& +0.85(8)~\cite{degroote2024AgPLB} & +0.880(35)\{3\} \\
\hline
\hline
\end{tabular}
\end{table}

\subsection{Nuclear electric quadrupole moments of Ag isotopes}
The quadrupole moment of the long-lived isomeric state $^{110m}$Ag is typically used as reference value~\cite{degroote2024AgPLB,vandenborne2025,berkes1984}. The $^{108m,110m}$Ag isomeric quadrupole moments have been extracted from hyperfine structure measurements of their $B$-factors in the $5p\,^2P^{o}_{3/2}$ atomic level using a simple phenomenological model to calculate the EFG~\cite{fischer1975}. Later, this EFG value was modified to take into account a Sternheimer correction factor~\cite{berkes1984}, leading to revised values for the quadrupole moments of $^{108m,110m}$Ag. The revised quadrupole moment of $^{110m}$Ag, $Q(^{110m}{\rm Ag})= +1.44(10)$\,$e$b, has typically been used as the reference moment since then; e.g. in the recent high-precision measurements by de Groote \textit{et al.}~\cite{degroote2024AgPLB} and van den Borne \textit{et al.}~\cite{vandenborne2025}, shown in Table~\ref{tab:Q_moments}.

To further improve the precision and accuracy of electric quadrupole moments across the Ag chain, the EFG has been calculated in this work for the $5p\,^2P^{o}_{3/2}$ and $4d^9\,5s^2\,^2D_{5/2}$ levels of the Ag atom. Table~\ref{TResultsEFG} shows the calculated value of the EFG constant $V_{zz}$ as a function of theoretical correction, where the total value can be combined with the experimental electric quadrupole hyperfine $B$-factor to extract the values of the electric quadrupole moments of Ag isotopes, as summarized in Table~\ref{tab:Q_moments}. Interestingly, in contrast to the EEs calculations, the EFG in the $4d^95s^2\,^2D_{5/2}$ state is quite insensitive to basis‐set corrections, particularly the higher‐order harmonics. However, this behavior is not unique: it is well-known that basis functions with higher orbital angular momentum $L$ are important to account for electron‐polarization effects that contribute to the EEs. Conversely, higher‐$L$ functions have reduced amplitude of the wave function in the nuclear region, leading to a suppressed direct contribution to the hyperfine constants, as confirmed by the results in Tables~\ref{TResults} and~\ref{TResultsEFG}.

\begin{table}
\caption{Calculated values of the electric field gradient $V_{zz}$ (in a.u.) for the $5p\,^2P^{o}_{3/2}$ 
and $4d^9\,5s^2\,^2D_{5/2}$ states
of Ag.}
\begin{tabular}{lll}
\toprule
\toprule
                    & $5p\,^2P^{o}_{3/2}$  &  $4d^9 5s^2\,^2 D_{5/2}$\\
\midrule                  
DHF                 &  +0.575              & $-$4.594       \\
CCSD(T)             &  +1.209              & $-$5.124       \\
CCSDT-3 $-$ CCSD(T) &  $-$0.003*            & +0.002(2)     \\
CCSDT $-$ CCSDT-3   &  +0.002*             & $-$0.001*  \\
CCSDT(Q) $-$ CCSDT  &  +0.003(3)           & $-$0.001(1)    \\
Basis-set correction&  +0.001(1)           &  $<10^{-3}$  \\
Breit               &  $-$0.006(3)          & +0.005(2)     \\
QED                 &  $-$0.001(1)          & $-$0.003(3)    \\
Total               & +1.204(4)            & $-$5.121(4)\\
Other studies      & +1.256(102)~\cite{vandenborne2025}    & $-$5.751(544)~\cite{dinger1989} \\
\hline
\hline
\end{tabular}
\label{TResultsEFG}

* Uncertainty is negligible within the numerical procedure.
\end{table}

\begin{figure}
    \centering
    \includegraphics[width=0.5\textwidth]{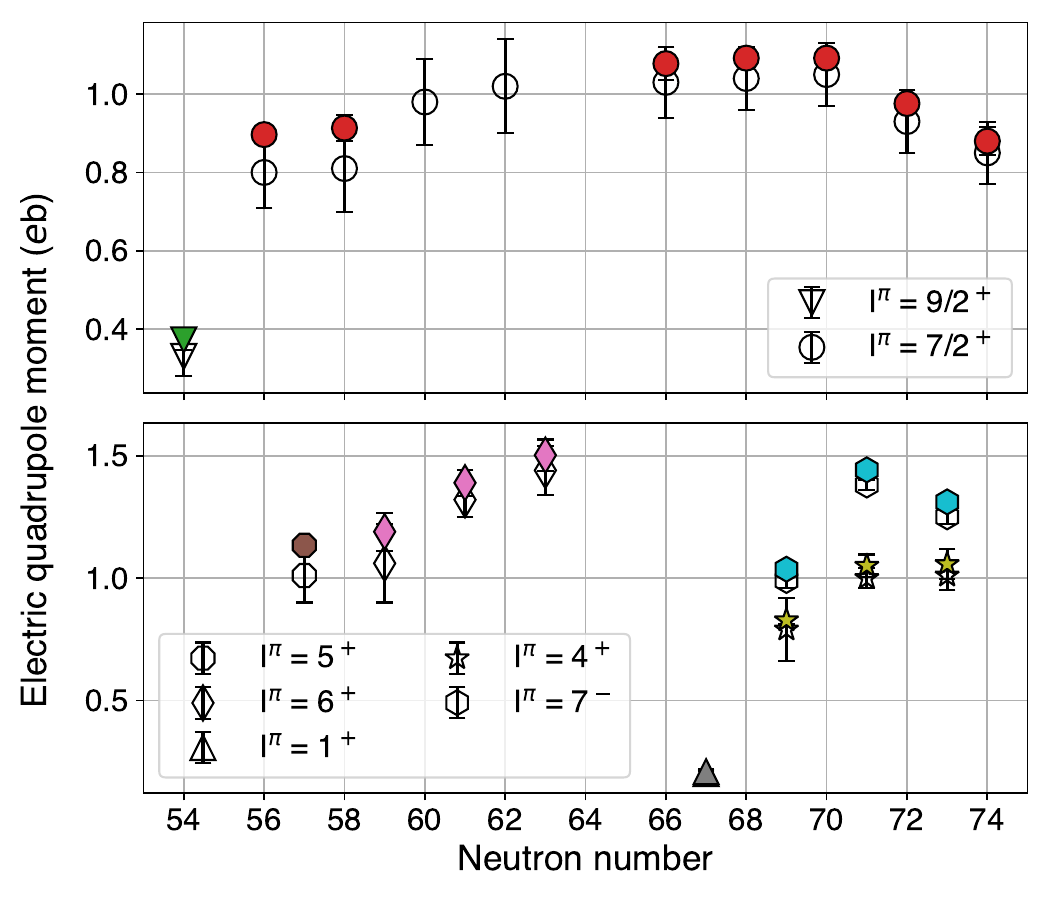}
    \caption{Overview of the literature values and the values determined in this work for the electric quadrupole moments in Ag, listed in Table~\ref{tab:Q_moments}. The empty markers correspond to the literature moments, and the filled markers to the moments from this work. The error bars represent the experimental uncertainty; the systematic uncertainty from the calculated EFG in this work is not shown, as it is smaller than the marker size. Only the quadrupole moments obtained from optical measurements are re-determined in this work, and the level-mixing values are shown unchanged for visual comparison.}
    \label{fig:Q_lit_new}
\end{figure}

One can see from Tables~\ref{TResults} and~\ref{TResultsEFG} that higher‑order correlation effects beyond the CCSD(T) level make only a modest contribution to the hyperfine constants of the $4d^95s^2\,^2D_{5/2}$ state. This is consistent with the overall small impact of correlation effects on hyperfine constants for this state. For example, including correlation effects at the CCSD(T) level alters the DHF value of $V_{zz}$ for this state by 11.5\%. This change is an order of magnitude smaller than in the case of the $5p\,^2P^{o}_{3/2}$ state, where the corresponding change reaches 110.3\%. The same holds for the $\Tilde{A}_0$ constants, with corresponding contributions of 47.9\%, 67.4\%, 89.8\%, and 5.5\% for the $5s\,^2S_{1/2}$, $5p\,^2P^{o}_{1/2}$, $5p\,^2P^{o}_{3/2}$, and $4d^95s^2\,^2D_{5/2}$ states, respectively.

For isotopes where the hyperfine structure of either the $5p\,^2P^{o}_{3/2}$ or $4d^95s^2\,^2D_{5/2}$ state has been reported in literature, the electric quadrupole moment can be readily extracted using Eq.~\eqref{eq:B} and $V_{zz}=+1.204(4)$~a.u. for $^2P^{o}_{3/2}$ or $V_{zz}=-5.121(4)$~a.u. for $^2D_{5/2}$, shown as $Q_{\rm{new}}$ in Table~\ref{tab:Q_moments}. 
Note that the error on the obtained quadrupole moments is now dominated by the experimental error on the measured hyperfine $B$-factors, while in the past it was dominated by the error on the estimated EFG.

Future spectroscopic measurements of the $5p\,^2P^{o}_{3/2}$ or $4d^95s^2\,^2D_{5/2}$ state for other isotopes can also be analyzed using our calculated values of the EFG. For measurements using other techniques or laser spectroscopy of other atomic/ionic states, Eq.~\eqref{eq:opticalQ} can be used along with measurements of a reference isotope whose hyperfine structure in the $5p\,^2P^{o}_{3/2}$ or $4d^95s^2\,^2D_{5/2}$ level has been reported, such as the common reference value $Q(^{110m}{\rm Ag})$, to extract $Q$ for other isotopes.

Figure~\ref{fig:Q_lit_new} shows the literature values (empty markers) and the values determined in this work (filled markers) for the electric quadrupole moments of Ag isotopes for ground and isomeric states. In addition to the significantly reduced error bars, the change in nominal value is also non-negligible for several nuclear states. Only the quadrupole moments of $^{107m,109m}$Ag are not determined here, as they were measured using level-mixing resonance in oriented nuclei rather than optical spectroscopy.

\section{Nuclear charge radii}
We finally turn our attention to nuclear charge radii. The evolution of the difference in mean-square charge radii across the Ag isotopic chain has been studied via laser spectroscopy in several past works, including by Dinger \textit{et al.}~\cite{dinger1989}, by Ferrer \textit{et al.}~\cite{ferrer2014}, and most extensively by Reponen \textit{et al.}~\cite{Reponen2021} with a partial re-evaluation in Ref.~\cite{vandenborne2025}. Both Ferrer \textit{et al.} and Reponen \textit{et al.} measured isotope shifts for the 328-nm $4d^{10}5s~^2S_{1/2}\rightarrow$~$4d^{10}5p~^2P_{3/2}$ transition, focusing on the neutron-deficient side and reaching down to $^{96}$Ag, while the earlier measurements by Dinger \textit{et al.}, also in neutron-deficient Ag, were reported for the 548-nm line~\cite{dinger1989}. Reponen \textit{et al.} measured up to $^{121}$Ag on the neutron-rich side, but a gap in the available data for nuclear ground states exists between $^{105}$Ag and $^{113}$Ag, with the exception of the stable $^{107,109}$Ag. 

In laser spectroscopy, the change in mean-square nuclear charge radii $\delta \langle r^2\rangle^{A',A}$ between two isotopes $A'$ and $A$ is related to the isotope shift $\delta \nu ^{A',A}$ in an optical transition frequency as
\begin{equation}
    \delta \nu ^{A',A} = F \delta \langle r^2 \rangle^{A',A} + K_{\rm{total}} \left(\frac{1}{M'}-\frac{1}{M}\right)
\end{equation}
where $M'$ and $M$ are the masses of isotopes $A'$ and $A$. The field-shift (FS) factor $F$ and the mass-shift factor $K_{\rm total}$ are electronic-structure parameters that are transition-dependent but isotope-independent. The mass-shift factor $K_{\rm{total}}=K_{\rm{NMS}}+K_{\rm{SMS}}$ is the sum of two contributions, the normal mass shift factor (NMS) and the specific mass shift (SMS) factor.

For elements that have three or more isotopes whose absolute charge radius has been measured, the electronic factors $F$ and $K_{\rm total}$ can be determined for any optical transition by relating the measured $ \delta \nu ^{A',A}$ with the known values of $\langle r^2 \rangle^{A',A}$ for these isotopes. In Ag, however, absolute charge radius measurements have been reported only for the two stable isotopes $^{107,109}$Ag. As a result, high-accuracy electronic-structure calculations of $F$ and $K_{\rm total}$ are needed to extract $\delta \langle r^2 \rangle^{A',A}$ from measured isotope shifts for any given transition.

\begin{figure}
    \centering
    \includegraphics[width=0.49\textwidth]{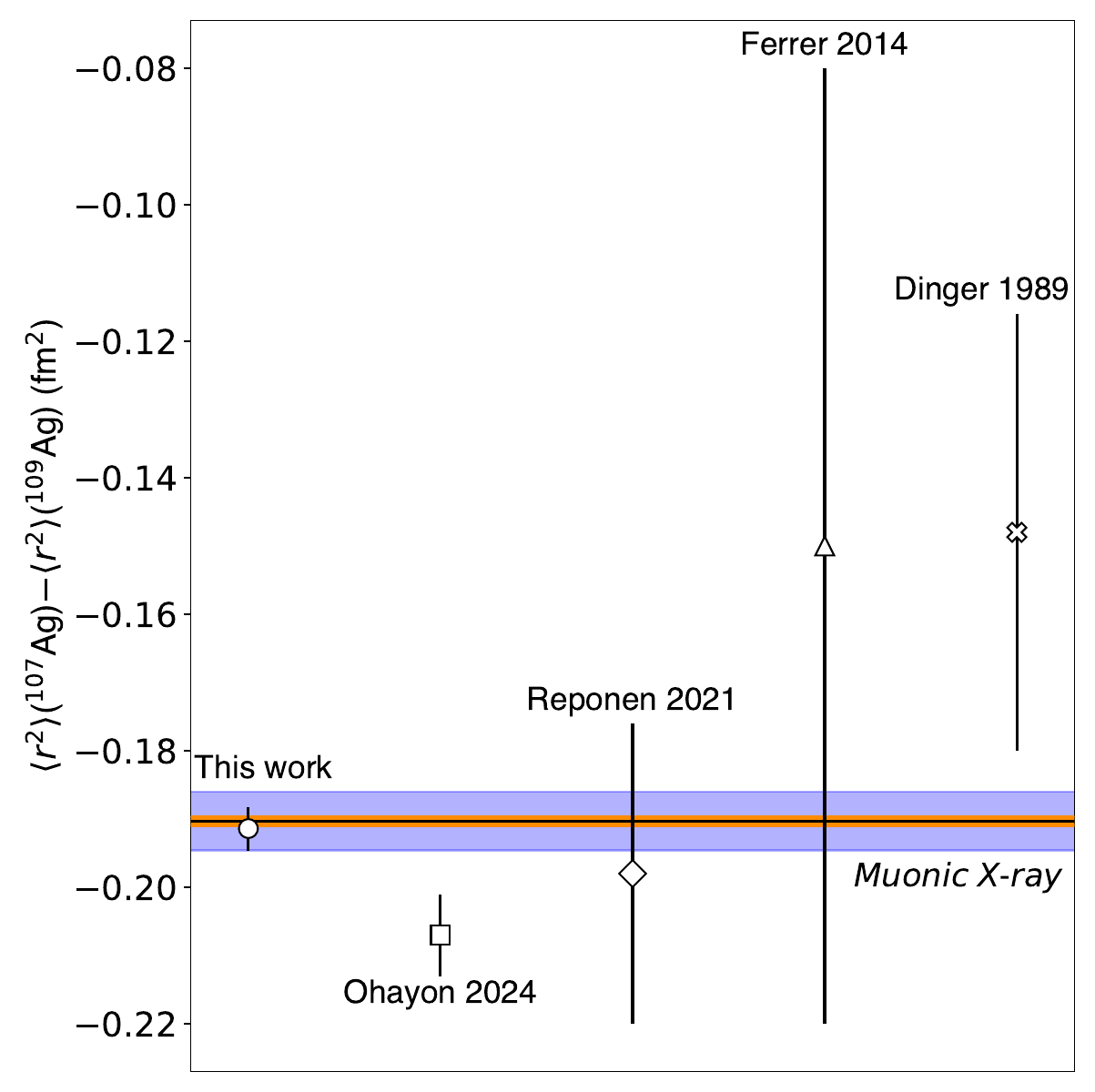}
    \caption{Comparison of the difference in mean-square nuclear charge radius between the stable isotopes $^{107,109}$Ag as reported in Refs.~\cite{Ohayon2024Ag,Reponen2021,ferrer2014,dinger1989} and this work, and the value extracted from the absolute charge radii derived from muonic X-ray spectroscopy~\cite{FrickeHeiligAg}, which is shown as a horizontal line. The purple band corresponds to the systematic uncertainty, and the orange to the statistical uncertainty of the value from muonic X-ray data.}
    \label{fig:stable}
\end{figure}

Such calculations had not been reported for the Ag atom until recently~\cite{Ohayon2024Ag}. Consequently, Refs.~\cite{dinger1989,ferrer2014,Reponen2021} used semi-empirical $F$ and $K_{\rm total}$ factors based on heuristic arguments regarding $ns$-$np$ atomic transitions and the ground-state hyperfine structure. Additionally, a scaling factor of $0.976$ was applied to $F$ to account for corrections due to higher-order radial moments.

The consequence of using semi-empirical $F$ and $K_{\rm total}$ factors was the limited final precision in the extracted values of $\delta \langle r^2 \rangle^{A',A}$, often with systematic errors that were unaccounted for. This is visualized in Fig.~\ref{fig:stable}, which compares the difference in the mean-square charge radii of $^{107,109}$Ag as reported by Dinger \textit{et al.}, Ferrer \textit{et al.}, and Reponen \textit{et al.}, and the value extracted from absolute charge radius measurements using muonic X-ray spectroscopy.

Figure~\ref{fig:stable} also shows the result by Ohayon \textit{et al.}~\cite{Ohayon2024Ag}, who recently performed the first \textit{ab initio} calculations of $F$ and $K_{\rm total}$ in Ag with analytical-response relativistic coupled-cluster (AR-RCC) theory including triple-excitation, basis-set, Breit, and QED corrections. Ohayon \textit{et al.} recommended a new set of mean-square nuclear charge radii for Ag and found that the semi-empirical $F$ and $K_{\rm total}$ factors that previous experimental campaigns used to extract $\delta \langle r^2 \rangle$ deviate significantly from the \textit{ab initio} results. While their recommended values of $\delta \langle r^2 \rangle^{A',A}$ have improved the agreement with nuclear density functional theory, Fig.~\ref{fig:stable} shows that their value for $\delta \langle r^2 \rangle^{107,109}$ disagrees with the muonic X-ray value by more than 2$\sigma$.

The limited accuracy and precision in the nuclear charge radii of Ag reported in literature calls for additional investigations. In this section, we report \textit{ab initio} coupled cluster calculations of $F$ and $K_{\rm total}$ for several optical transitions in Ag that can be used to re-analyze isotope shifts reported in literature, as well as future experiments.

\subsection{Theory of atomic isotope-shift factors}
In the present work, the FS constant $F$ is defined as 
\begin{equation}
\label{FSdef}
F= \frac{d\, \nu}{d \left\langle r^2 \right\rangle} , 
\end{equation}
where 
the derivative is calculated at the point $r_{\rm rms}=\sqrt{\langle r^2 \rangle}=4.564$~fm, which corresponds to the reference isotope $^{109}$Ag~\cite{ANGELI201369}.

In Ref.~\cite{skripnikov2024isotopeQED}, a model approach to treat the contribution of QED effects to the FS constant in many-electron systems was introduced. In this approach, one starts from the QED contribution to the electronic energy as a function of $\langle r^2 \rangle$:
\begin{equation}
E^{\rm QED} \left( \langle r^2 \rangle \right) = \langle \Psi^{\langle r^2 \rangle} | \sum_i H^{\rm QED}(\mathbf{r}_i, \langle r^2 \rangle) | \Psi^{\langle r^2 \rangle} \rangle,   
\label{EQED}
\end{equation}
where $|\Psi^{\langle r^2 \rangle} \rangle$ is the wave function of the many-electron atom, which depends on the nuclear charge radius through the electron-nucleus interaction operator.
$H^{\rm QED}$ is given by the sum of vacuum-polarization (VP) and self-energy (SE) terms:
\begin{equation}
    H^{\rm QED} = H^{\rm VP}+H^{\rm SE}.
    \label{eq:h_se_vp}
\end{equation}
Formally, one can then calculate the QED contribution to the transition frequency~$\nu$ using Eq.~\eqref{FSdef}. For the VP contribution, $H^{\rm VP}(\mathbf{r}, \langle r^2 \rangle)$ can be well approximated by the Uehling potential as a known function of $\langle r^2 \rangle$. However, the calculation of the SE contribution is not as straightforward. In Ref.~\cite{skripnikov2024isotopeQED}, a generalization of the model QED approach~\cite{Shabaev:13} was introduced to calculate this contribution using the following expression:
\begin{eqnarray}\label{Fse}
    F^{\rm SE} = \sum_{p,q} 
\left.\frac{d X_{p,q}^{\langle r^2 \rangle}}
    {d \langle r^2 \rangle}\right|_{\langle r_0^2 \rangle} D_{p,q}^{\langle r_0^2 \rangle} 
    +\sum_{p,q} 
    X_{p,q}^{\langle r_0^2 \rangle}
 \left.\frac{d D_{p,q}^{\langle r^2 \rangle}}{d \langle r^2 \rangle}\right|_{\langle r_0^2 \rangle}. 
\end{eqnarray}
Here, $D_{p,q}^{\langle r^2 \rangle}$ is the correlated one-electron density matrix written in the basis of functions $h_{p}(\mathbf{r})=h_{kljm}(\mathbf{r})$ (\textit{p} is the multi-index), which are linear combinations of functions $\widetilde{h}_{nljm}(\mathbf{r})=\eta_{nljm}(\mathbf{r}) \theta(R_{\rm cut}-|\mathbf{r}|)$. The functions $\eta_{nljm}(\mathbf{r})$ correspond to an H-like ion of a given element (with a principal quantum number $n \le 5$). The function $\theta(R_{\rm cut}-|\mathbf{r}|)$ is the Heaviside step function, and $R_{\rm cut}$ is a small radius that can be varied slightly to study the stability of the results obtained~\cite{Skripnikov:2021a}. $\langle r_0^2 \rangle$ is the mean-square nuclear charge radius corresponding to the reference isotope, and all derivatives are taken at this point. $X_{p,q}=X_{kljm,k'ljm}$ in Eq.~\eqref{Fse} are the matrix elements of the SE operator over the $h_{p}$ functions, and thus, to a good accuracy, they are linear combinations of the corresponding matrix elements over the H-like functions. The derivative $d X_{p,q}^{\langle r^2 \rangle} / d \langle r^2 \rangle$ can be calculated~\cite{skripnikov2024isotopeQED} using the analytic dependence of SE matrix elements on the finite nuclear size, as studied in Refs.~\cite{Milstein:2002,Milstein:2003,Milstein:2004,Yerokhin:2011b,Ulrich:2003,Pachucki:1993,ShabaevNS:1993,Eides:1997}.
Note that it is not necessary to calculate the density matrix in Eq.~\eqref{Fse} explicitly, and the finite-field technique can be employed.

To the first order in the electron-to-nucleus mass ratio $m/M$, one can calculate the mass shift effect using the following relativistic operators within the Breit approximation~\cite{shabaev1985, shabaev1988, palmer1987reformulation, shabaev1994rel}:
\begin{equation}
\label{opNMS}
    H_{\rm NMS} = \frac{1}{2M}\sum_i\left(\mathbf{p}_i^2-\frac{\alpha Z}{r_i}\left[ \bm{\alpha}_i+\frac{(\bm{\alpha}_i\cdot \mathbf{r}_i)\mathbf{r}_i}{r_i^2}\right]\cdot\mathbf{p}_i\right),
 \end{equation}
\begin{equation}
\label{opSMS}
    H_{\rm SMS} = \frac{1}{2M}\sum_{i \neq k}\left(\mathbf{p}_i\cdot\mathbf{p}_k-\frac{\alpha Z}{r_i}\left[ \bm{\alpha}_i+\frac{(\bm{\alpha}_i\cdot \mathbf{r}_i)\mathbf{r}_i}{r_i^2}\right]\cdot\mathbf{p}_k\right),
\end{equation}
where $Z$ is the proton number. The constants $K_{\rm NMS}$ and $K_{\rm SMS}$ (up to a multiplier of $1/M$) can be calculated as expectation values of these operators [Eqs.~\eqref{opNMS}, \eqref{opSMS}] with the wave functions obtained using the Dirac-Coulomb-Breit Hamiltonian.

Operator~\eqref{opSMS} is a two-electron operator. Therefore, if some inner-core electrons are excluded from the correlation calculation of a many-electron atom, the effective one-electron operator that accounts for the core-valence SMS interaction must still be considered even to properly calculate corrections due to correlation effects.

For highly-charged ions, Eq.~\eqref{opNMS} gives terms of order $(\alpha Z)^4m^2/M$ and zero order in $\alpha$ for the MS. Terms of order $(\alpha Z)^5m^2/M$ and higher orders in $\alpha Z$ can be calculated within the rigorous QED theory developed in Refs.~\cite{shabaev1985, shabaev1988, ShabaevRecoil:98}; see also Refs.~\cite{Pachucki:1995:1854, yelkhovsky1994, Adkins:2007:042508}. However, a direct application of such theory to neutral many-electron atoms is complicated. In Ref.~\cite{Anisimova:2022}, the model QED approach was developed to account for these high-order QED corrections to the mass shift~\cite{Anisimova:2022} beyond the approximation [Eq.~\eqref{opNMS}]. In Ref.~\cite{skripnikov2024isotopeQED}, this approach was adapted for application to many-electron systems within the relativistic coupled cluster methods.

An additional QED contribution to the first order in $\alpha$ to the nuclear recoil effect arises from the perturbation of the electronic wave function by the VP and SE interactions~\cite{King:2022:43, Sahoo:2021}. This contribution can be estimated by calculating the difference between the values of the mass-shift constants determined with and without the inclusion of the Uehling and model SE operators in the electron correlation calculation. Thus, the same approximation for the mass shift operators [Eqs.~\eqref{opNMS}, \eqref{opSMS}] is employed. It should be noted that, in the case of the neutral Al atom in Ref.~\cite{skripnikov2024isotopeQED}, this contribution was shown to be negligible compared to the QED contribution described above.

\subsection{Computational results}

\begin{table*}
\centering
\caption{Values of the IS atomic factors for Ag obtained in this work (TW) and Refs.~\cite{Reponen2021,Ohayon2024Ag}.}
\begin{tabular}{lrrrrr}
\toprule
\toprule
& $5s\,^2S_{1/2} \to  5p\,^2P^{o}_{1/2}$ & $5s\,^2S_{1/2} \to  5p\,^2P^{o}_{3/2}$ & $5s\,^2S_{1/2} \to  6s\,^2S_{1/2}$ & $5s\,^2S_{1/2} \to  6p\,^2P^{o}_{1/2}$ & $5s\,^2S_{1/2} \to  6p\,^2P^{o}_{3/2}$ \\
$\lambda$  &  338~nm   &   328~nm &  235~nm &  207~nm  & 206~nm  \\
\hline
\\
\multicolumn{6}{c}{ $F$, MHz/fm$^2$ }   \\
\hline
DHF                    & $-$2838     & $-$2858   & $-$2431   & $-$2880  & $-$2886    \\
CCSD(T)                & $-$3676     & $-$3712   & $-$3351   & $-$3826  & $-$3831    \\
CCSDT-3 $-$ CCSD(T)    & 6(6)      & 11(11)  & $-$10(10) & 2(2)   & 7(7)     \\
CCSDT $-$ CCSDT-3      & 15(15)    & 15(15)  & 14(14)  & 17(17) & 18(18)   \\
CCSDT(Q) $-$ CCSDT     & 16(16)    & 16(16)  & 16(16)  & 16(16) & 16(16)   \\
Breit                  & 14(7)     & 14(7)   & 13(7)   & 15(7)  & 15(7)    \\
Basis set corr.*       & 0(33)     & 0(33)   & 0(30)   & 0(33)  & 0(33)    \\
Nuc. model             & $-$18(9)    & $-$18(9)  & $-$17(8)   & $-$19(9) & $-$19(9)   \\
QED-VP                 & $-$20(6)    & $-$21(6) & $-$19(6)  & $-$21(6) & $-$21(6)    \\
QED-SE                 & 39(12)    & 39(12)  & 36(11)  & 40(12) & 40(12)   \\
Total, TW              & $-$3624(44) & $-$3655(45) & $-$3317(41) & $-$3776(44)  & $-$3776(45)    \\ 
Ref.~\cite{Ohayon2024Ag}  & $-$3525(48)  & $-$3557(49) & $-$3223(36) & $-$3680(43)  & $-$3683(43)   \\      
Ref.~\cite{Reponen2021}   &           & $-$4300(300) & & &  \\ 

\\
\multicolumn{6}{c}{$K_{\rm NMS}$, GHz u}   \\
\hline
CCSD(T)                  & $-$475     & $-$489     & $-$684    & $-$778    & $-$781    \\
CCSDT-3 $-$ CCSD(T)      & $-$6(2)    & $-$7(2)    & $-$10(3)  & $-$11(3)  & $-$11(3)  \\
CCSDT $-$ CCSDT-3    & 2(2)     & 2(2)     & 3(3)    & 4(4)    & 4(4)    \\
CCSDT(Q) $-$ CCSDT   & 2(2)     & 2(2)     & 4(4)    & 3(3)    & 3(3)    \\
Breit                    & 1(0)     & 1(1)     & 2(1)    & 2(1)    & 2(1)    \\
Total (non-QED), TW     & $-$476(3)  & $-$490(4)  & $-$686(7) & $-$780(7) & $-$784(7) \\
QED **                    & $-$92(28)  & $-$94(28)  & $-$85(25) & $-$97(29) & $-$97(29) \\
Total, TW         & $-$568(28) & $-$583(28) & $-$771(26)& $-$877(30)& $-$880(30)\\
Ref.~\cite{Ohayon2024Ag}  & $-$431(5)  & $-$448(4) & $-$575(10) & $-$665(9)  & $-$671(9)   \\    
\\
\multicolumn{6}{c}{$K_{\rm SMS}$, GHz u}   \\
\hline
CCSD(T)                   & $-$711      & $-$712     & $-$775 & $-$858 & $-$857 \\
CCSDT-3 $-$ CCSD(T)       & $-$64(19)   & $-$64(19)  & $-$99(30)      & $-$98(29)     & $-$97(29)      \\
CCSDT $-$ CCSDT-3        & 10(10)    &  11(11)  &  9(9)    &  12(12) & 12(12)      \\
CCSDT(Q) $-$ CCSDT       & 28(28)    &  28(28)  &  53(53)  & 52(52)  & 52(52)   \\
Basis set corr.           & 5(5)      & 6(6)     & 8(8)     & 9(9)    & 9(9)     \\
Breit                     & 1(0)      & $-$1(0)    & $-$1(1)    & $-$1(0)   & $-$1(1)    \\
QED                       & $-$3(3)     & $-$3(3)    & $-$2(2)    & $-$2(2)   & $-$2(2)    \\
Total, TW          &  $-$734(36) & $-$734(36)  & $-$806(62)  &  $-$886(62)  & $-$884(61) \\
Ref.~\cite{Ohayon2024Ag}  & $-$1058(12) & $-$1031(13) & $-$1260(22) & $-$1375(23)  & $-$1363(23)   \\    
\\
\multicolumn{6}{c}{$K_{\rm NMS}+K_{\rm SMS}$, GHz u}                             \\
Total, TW          & $-$1302(45) & $-$1318(46) &  $-$1577(68) &  $-$1763(69) & $-$1765(68)           \\
Ref.~\cite{Reponen2021}   &           & $-$1956(360) & & &  \\
Ref.~\cite{Ohayon2024Ag}  & $-$1489(13) & $-$1479(14) & $-$1835(24) & $-$2040(25)  & $-$2034(25)   \\    
\hline
\hline
\end{tabular}
\begin{flushleft}
* Only the uncertainty was estimated; see text.

**  This QED contribution includes terms that were not considered in previous theoretical studies of IS in neutral Ag.
\end{flushleft}

\label{TabISFactors}
\end{table*}
Table~\ref{TabISFactors} provides the calculated values of the $F$ and $K$ factors for the 338-, 328-, 235-, 207-, and 206-nm transitions in Ag. As can be seen from Table~\ref{TabISFactors}, the computational scheme exhibits smooth convergence with respect to higher-order correlation effects. The majority of correlation effects are accounted for at the all-electron CCSD(T) level, with a non-iterative treatment of triple excitation amplitudes. Corrections for iterative triple excitations are relatively small. Finally, while the contributions from quadruple excitations are small, they are not entirely negligible and give the dominant contribution to the uncertainty of the MS constant.
As an additional verification, we compared the values of the $F$ constants calculated at the DHF level (corresponding to the Ag$^+$ ion) using the {\sc dirac} code, which employs Gaussian-type basis functions, with those obtained using the atomic finite-difference {\sc hfd} code~\cite{Bratzev:77,HFD}, and found only a negligible deviation of less than 0.016\% between the results obtained from these two completely different and independent implementations of the Dirac-Hartree-Fock method.

\begin{table*}
    \centering
    \caption{Difference in mean-square nuclear charge radii of the stable Ag isotopes $\delta \langle r^2 \rangle  ^{107,109}$ extracted from optical IS measurements $\delta \nu^{107,109}$ and other experiments. The label 'TW' stands for the results of this work. The error on the muonic radius labeled 'Nuc.\,pol.' corresponds to the uncertainty due to nuclear polarization effects.}
    \label{tab107109}
    \begin{threeparttable}
    \begin{tabular}{lll}
        \toprule
        \toprule
        Transition & $\delta \nu_{107,109}$ (MHz) & $\delta r^2_{107,109}$ (fm$^2$) \\
        \hline
$5s \, ^2S_{1/2} \to 5p \, ^2P_{1/2}$ \hspace{1cm} & 473(4)~\cite{Fischer1968,Badr2004} \hspace{1.5cm} & \begin{tabular}{@{}c@{}}$-$0.2067(11)$_{\text{exp}}$(29)$_{K,F}$~\cite{Ohayon2024Ag} \\ 
$-$0.1922(11)$_{\text{exp}}$(32)$_{K,F}$, TW \end{tabular} \\ \hline
$5s \, ^2S_{1/2} \to 5p \, ^2P_{3/2}$ \hspace{1cm} & 473.2(7)~\cite{Ohayon2024Ag} & \begin{tabular}{@{}c@{}}$-$0.2044(2)$_{\text{exp}}$(29)$_{K,F}$~\cite{Ohayon2024Ag} \\ 
$-$0.1914(2)$_{\text{exp}}$(32)$_{K,F}$, TW \end{tabular} \\\hline
$5s \, ^2S_{1/2} \to 6s \, ^2S_{1/2}$ \hspace{1cm} & 368(7)~\cite{Ohayon2024Ag,Fischer1968,Badr2004} & \begin{tabular}{@{}c@{}}$-$0.2120(22)$_{\text{exp}}$(27)$_{K,F}$~\cite{Ohayon2024Ag} \\ 
$-$0.1926(21)$_{\text{exp}}$(43)$_{K,F}$, TW \end{tabular} \\\hline
$5s \, ^2S_{1/2} \to 6p \, ^2P_{3/2}$ \hspace{1cm} & 414.5(6)~\cite{Badr:2006,Elbel1962} & \begin{tabular}{@{}c@{}}$-$0.2073(2)$_{\text{exp}}$(27)$_{K,F}$~\cite{Ohayon2024Ag} \\ 
$-$0.1900(2)$_{\text{exp}}$(39)$_{K,F}$, TW \end{tabular} \\\hline
Muonic Ag~\cite{FrickeHeiligAg} & & $\bullet$$-$0.1903(7)$_{\text{exp}}$(43)$_{\text{nuc.pol.}}$\\
        \hline
        \hline
    \end{tabular}
\begin{tablenotes}[]
\item[] ($\bullet$) Calculated from the difference in Barrett radii and V$_2$ values of $^{107,109}$Ag~\cite{FrickeHeiligAg}.
\end{tablenotes}
\end{threeparttable}
\end{table*}

The calculation of $K_{\rm SMS}$ constants was among the most challenging in this study. As one can see from Table~\ref{TabISFactors}, electron correlation effects beyond the CCSD(T) model contribute more significantly to SMS constants than to the FS and NMS ones. Therefore, additional test calculations were conducted specifically for the SMS constants to independently verify them. All calculations in our computational scheme utilized single-reference coupled cluster (SR-CC) methods. To cross-check the results, we employed a markedly different CC model, the Fock-space coupled cluster (FS-CC) method~\cite{Lindgren:87,Kaldor:91,Visscher:01,Musial:04,Eliav:Review:22} in its relativistic implementation~\cite{EXPT_website,Oleynichenko_EXPT,Zaitsevskii:2023,Oleynichenko:CCSDT:20}. 
As with the SR-CCSDT case, the FS-CC calculations including single, double, and full triple excitations (FS-CCSDT) did not practically allow for the inclusion of all Ag electrons in the correlation treatment due to the high computational cost associated with 47 electrons. In the relativistic FS-CCSDT model used~\cite{Oleynichenko:CCSDT:20}, all outer-core and valence electrons included in the correlation calculation are treated on an equal footing just as in the SR-CCSDT model. Therefore, for the Ag atom in the electronic states with single-reference character considered here, the quality of these two models is expected to be similar. Indeed, according to our calculations, the $K_{\rm SMS}$ constants obtained in the 29-electron
FS-CCSDT model within the SBas basis set (see Appendix for details) deviate from those in the 29-electron SR-CCSDT model by approximately 20~GHz\,u for $5s~^2S_{1/2} \to 5p~^2P^{o}_{1/2}$ and $5s~^2S_{1/2} \to 5p~^2P^{o}_{3/2}$ transitions, and by about 40~GHz\,u for all other transitions involving the $6s$ or $6p$ states, which is within our final estimated uncertainty (Table~\ref{TabISFactors}).  This cross-check using significantly different models (SR-CC and FS-CC) and codes provides a strong additional verification of the present results. 

Notably, the inclusion of fully treated triple excitations for 29 correlated electrons with the Fock-space method, i.e., the difference between the FS-CCSDT and FS-CCSD values, led to a decrease in the absolute values of the $K_{\rm SMS}$ constants by approximately 100~GHz\,u for the $5s~^2S_{1/2} \to 5p~^2P^{o}_{1/2}$ and $5s~^2S_{1/2} \to 5p~^2P^{o}_{3/2}$ transitions, and by approximately 200~GHz\,u for all other transitions involving the $6s$ or $6p$ states of Ag. This demonstrates a significant contribution of triple excitations. Interestingly, for 19 correlated electrons, the contribution of triple excitations within the FS-CC model is slightly lower and amounts to about 50~GHz\,u for the $5s\,^2S_{1/2} \to 5p\,^2P^{o}_{1/2}$ and $5s\,^2S_{1/2} \to 5p\,^2P^{o}_{3/2}$ transitions, and approximately 125~GHz\,u for transitions involving the $6s$ or $6p$ states of Ag. This result reveals that the role of triple excitations in the FS-CC approach is significant for the ``core'' $3d^{10}$ electrons of Ag. This finding partly motivated us to adopt the state-specific SR-CCSD(T) method as our primary model for the Ag atom, in which the leading effects of triple excitations are taken into account for all 47 electrons.

One should emphasize the large contribution of $11\%$ to $16\%$ due to QED effects to the NMS factors for all considered transitions. As was mentioned above, this QED contribution consists of two components. The first one is the contribution beyond the approximation [Eq.~\eqref{opNMS}] and is of zero order in the fine-structure constant $\alpha$. The second one can be considered as the interference of the VP and SE model potentials with the interaction given by Eq.~\eqref{opNMS}, and is of the first order in $\alpha$. According to our calculations, the first QED component is 40 times larger than the second one. Therefore, the latter can actually be neglected for the present study. 
As far as we know, the first QED contribution has not been previously considered in Ag.
Only recently, consideration of this component became possible due to the introduction of the model nuclear recoil QED approach suggested in Ref.~\cite{Anisimova:2022} and adapted for correlation calculations in Ref.~\cite{skripnikov2024isotopeQED}. Therefore, one should use the values labeled ``Total (non-QED)'' in Table~\ref{TabISFactors} to compare with previous theoretical estimations of the NMS atomic factor. Note that the QED contribution is larger than the overall uncertainty of the MS constant and thus must be taken into account at the present level of accuracy.

The QED contribution to the FS constant is small but not negligible. The SE contribution to the FS constant amounts to 1\% of the total FS constant value. This is about five times larger than the corresponding SE contribution to the FS factor in Al~\cite{skripnikov2024isotopeQED} demonstrating a significant atomic-number dependence.

The contribution due to the Breit interaction is small for all IS constants. For both the total MS and FS, it is smaller than the contribution due to QED effects.

Table~\ref{TabISFactors} also compares the $F$ and $K$ factors calculated in this work with the values presented by Ohayon \textit{et al.}~\cite{Ohayon2024Ag}, also from \textit{ab initio} electronic structure calculations, and the factors used by Reponen \textit{et al.} most recently~\cite{Reponen2021}. Expectedly, both the values from this work and from Ohayon \textit{et al.} differ significantly from the semi-empirical values used in prior studies.

For the $F$ factor, the present values are in 2$\sigma$ agreement with those by Ohayon \textit{et al.} On the other hand, the agreement for $K$ factor is poor, especially for $K_{\rm SMS}$. Considering that in both the present work and the work of Ohayon \textit{et al.} the factors were calculated with relativistic coupled cluster theory (but different versions of RCC and different codes) and including higher-order corrections, it is unclear what the source of the limited agreement is.

\subsection{Nuclear charge radii of Ag isotopes}
Table~\ref{tab107109} provides the values of $\delta \langle r^2 \rangle^{107,109}$ deduced from the IS measurements for four transitions in Ag, compared with previous results obtained using different approaches. It can be seen that, for all considered transitions, our IS factors yield $\delta \langle r^2 \rangle^{107,109}$ values that are close to each other with a small spread and in 1$\sigma$ agreement with the value from muonic data~\cite{FrickeHeiligAg}. Such agreement gives confidence in the present results.

The difference in mean-square charge radii between the stable isotopes $^{107,109}$Ag, as reported in the different works, can be compared with the value determined from muonic X-ray spectroscopy of stable Ag. Fricke and Heilig~\cite{FrickeHeiligAg} conducted a combined analysis of muonic X-ray, electron scattering, and optical spectroscopy data and recommended $\langle r^2 \rangle ^{1/2}$($^{107}$Ag$)=4.544$~fm and $\langle r^2 \rangle ^{1/2}$($^{109}$Ag$)=4.565$~fm, without reporting an uncertainty, citing the low quality of electron scattering experiments that define the nuclear distribution model used in the combined analysis. Due to the lack of uncertainties in the absolute charge radii, the value of $\delta \langle r^2 \rangle^{107,109}=-0.1903(7)_{\text{exp}}$(43)$_{\text{nuc.pol.}}$~fm$^2$ is extracted from the difference in Barrett radii $R^{A}_k$ and the $V^A_2$ values from muonic-atom X-ray spectroscopy in Ref.~\cite{FrickeHeiligAg}, as $\delta \langle r^2 \rangle^{107,109} = (R^{107}_k / V^{107}_2)^2 - (R^{109}_k / V^{109}_2)^2$. The statistical and systematic uncertainty originates from the measured muonic X-ray energy and the uncertainty on the calculated nuclear polarization, respectively. We note that Ohayon \textit{et al.} also derived $\delta \langle r^2 \rangle^{107,109}$ with the same approach, but we are unable to reproduce the value of $-0.198(0)_{\rm exp}(4)_{\text{nuc.pol.}}(5)_{\text{nuc.shape}}$ that they reported~\cite{Ohayon2024Ag} and the origin of the nuclear-shape uncertainty they attached to the value, despite both works using the Fricke and Heilig analysis as the basis~\cite{FrickeHeiligAg}.

The values of $\delta \langle r^2 \rangle^{107,109}$ that are reported in different works, including the present one, are compared against it in Fig.~\ref{fig:stable}. Our result agrees with Reponen \textit{et al.}~\cite{Reponen2021}, Ferrer \textit{et al.}~\cite{ferrer2014}, and the muonic X-ray value within 1$\sigma$, with Dinger \textit{et al.}~\cite{dinger1989} within 2$\sigma$, while differing from Ohayon \textit{et al.}~\cite{Ohayon2024Ag} by approximately 3$\sigma$ for the 328-nm transition.

The 3$\sigma$ difference with Ohayon \textit{et al.} for the 328-nm transition is of importance because both works present relativistic coupled cluster calculations of the Ag atom. As seen in Table~\ref{tab107109}, the difference between our work and Ohayon \textit{et al.} is 3.0-3.8$\sigma$ across the different optical transitions, neglecting the common experimental error. Simultaneously, the agreement between this work and the value from muonic data is within 0.4$\sigma$ for all transitions, while between Ohayon \textit{et al.} and the muonic value it is 2.7-3.9$\sigma$. Since the same experimental isotope shifts underlie the charge radius differences in both this work and Ref.~\cite{Ohayon2024Ag}, this discrepancy primarily reflects differences in the calculated atomic isotope-shift factors. The agreement of the present results with the independent muonic-data determination therefore provides additional support for the isotope-shift factors calculated in this work, although the origin of the discrepancy with Ref.~\cite{Ohayon2024Ag} remains to be understood.

\begin{figure*}
    \centering
    \includegraphics[width=0.8\textwidth]{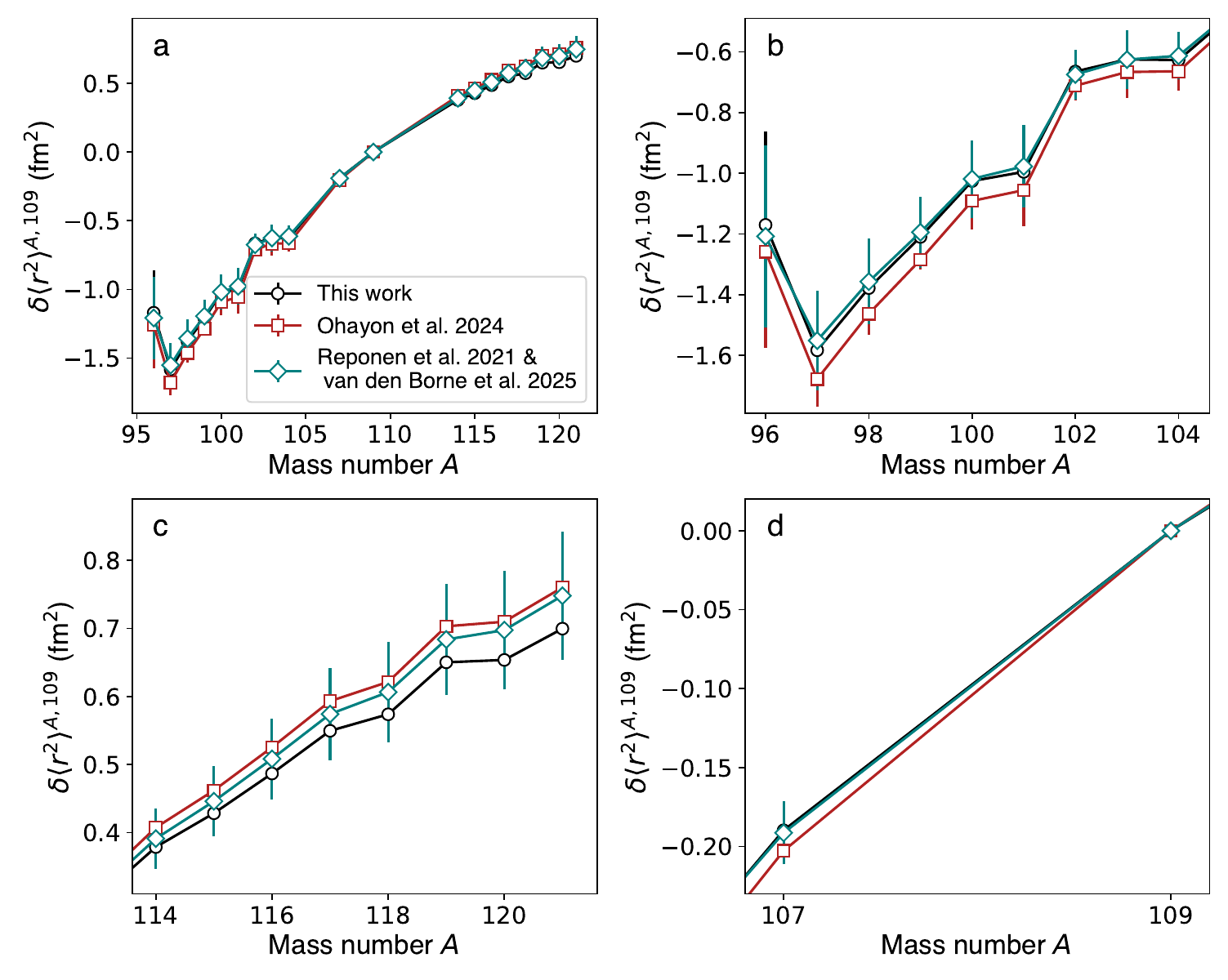}
    \caption{{(a)} Changes in mean-square nuclear charge radii of the ground states of the Ag chain (except for $A=113$ where $^{113m}$Ag is used), as extracted using the isotope shifts $\delta \nu^{A,109}$ reported in Refs.~\cite{Reponen2021,vandenborne2025} and the isotope-shift factors $F$ and $K$ calculated in this work (black circles) and by Ohayon \textit{et al.}~\cite{Ohayon2024Ag} (red squares). The values originally reported by Reponen \textit{et al.}~\cite{Reponen2021} and van den Borne \textit{et al.}~\cite{vandenborne2025} are also shown (teal diamonds). The error bars include both systematic and statistical errors. {(b)-(d)} Zoomed views of the neutron-deficient, neutron-rich, and stable regions of the chain, respectively.}
    \label{fig:dr2}
\end{figure*}

We choose the set of isotope shifts for the $5s\,^2S_{1/2} \to  5p\,^2P^{o}_{3/2}$ 328-nm transition reported by Reponen \textit{et al.}~\cite{Reponen2021} to extract revised mean-square nuclear charge radii using the $F$ and $K$ factors calculated in this work. The resulting values for $\delta \langle r^2 \rangle^{A,109}$ are extracted and shown in Fig.~\ref{fig:dr2}, where they are compared with $\delta \langle r^2 \rangle^{A,109}$ as reported by Reponen \textit{et al.}~\cite{Reponen2021} and van den Borne \textit{et al.}~\cite{vandenborne2025}, and also as extracted using the $F$ and $K$ constants calculated by Ohayon \textit{et al.} The asymmetric errors in $\delta \nu^{A,109}$ reported in Ref.~\cite{Reponen2021} were taken to be symmetric using the larger of the partial errors, to facilitate the error propagation.

\begin{table}[]
\caption{Revised values for the change in mean-square nuclear charge radii $\delta \langle r^2 \rangle ^{A,109}$ in Ag isotopes with respect to $^{109}$Ag with their nuclear spin $I$, and absolute charge radii $R_{\rm{ch}}$, using the calculated $F$ and $K$ from this work and the 328-nm isotope shifts reported by Ferrer \textit{et al.}~\cite{ferrer2014}, Reponen \textit{et al.}~\cite{Reponen2021}, van den Borne \textit{et al.}~\cite{vandenborne2025}, except for $^{107}$Ag, where the isotope shift from Ref.~\cite{Ohayon2024Ag} is used. Errors in parentheses are statistical from the isotope shifts, and in curly brackets are systematic from the isotope-shift factors. The uncertainties for $R_{\rm{ch}}$ include only those propagated from the isotope-shift factors and experimental error bars and do not include the unknown common uncertainty of the absolute radius for $^{109}$Ag used as the reference value.} \label{table:dr2}
\begin{tabular}{llll}
\toprule
\toprule
Isotope\hspace{0.4cm}   & $I$\hspace{0.8cm} & $\delta \langle r^2 \rangle ^{A,109}$~(fm$^2$)\hspace{1cm}    & $R_{\rm{ch}}$~(fm)         \\\hline
96  & (8)   &$-$1.17(31)\{2\}   & 4.435(35)\{2\}   \\
97  & (9/2) &$-$1.58(9)\{2\}    & 4.388(10)\{3\}  \\
98  & (5)   &$-$1.38(7)\{2\}    & 4.412(8)\{2\}   \\
99  & (9/2) &$-$1.21(3)\{2\}    & 4.431(3)\{2\}   \\
99m & (1/2) &$-$1.19(27)\{2\}   & 4.433(30)\{2\}  \\
100 & (5)   &$-$1.03(9)\{2\} & 4.451(10)\{2\}   \\
101 & 9/2   &$-$1.00(11)\{2\} & 4.455(13)\{2\}    \\
101m& (1/2) &$-$0.92(6)\{2\} & 4.463(6)\{2\}  \\
102 & 5     &$-$0.665(47)\{11\} & 4.4916(52)\{13\}   \\
103 & 7/2   &$-$0.625(82)\{10\} & 4.4960(91)\{11\}   \\
104 & 5     &$-$0.627(60)\{10\} & 4.4958(67)\{11\}   \\
107 & 1/2 &$-$0.1914(2)\{32\} & 4.5440(1)\{4\} \\
109 & 1/2 &0.0(0)\{0\}        & 4.565 \\
113m& 7/2 &0.2952(14)\{55\}  & 4.5972(2)\{6\} \\
114 & 1   & 0.379(5)\{7\}       & 4.6063(6)\{7\} \\
115 & 1/2 & 0.4285(14)\{80\}  & 4.6117(2)\{9\} \\
115m& 7/2 & 0.416(2)\{8\}       & 4.6103(2)\{9\} \\
116 &  1  & 0.487(3)\{9\}       & 4.6180(3)\{10\}  \\
116m1& 4  & 0.459(8)\{9\} & 4.6150(9)\{10\} \\
116m2& 7  & 0.455(8)\{9\} & 4.6146(9)\{10\} \\
117 & 1/2 & 0.550(8)\{10\} & 4.6248(8)\{11\} \\
117m& 7/2 &0.521(5)\{10\}      & 4.6218(5)\{11\} \\
118 & 4   & 0.574(5)\{11\}      & 4.6274(6)\{12\}  \\
118m1& 0  & 0.588(4)\{11\} & 4.6289(4)\{12\} \\
118m2& 7  & 0.568(4)\{11\} & 4.6268(4)\{12\} \\
119 & 1/2 &0.650(4)\{13\} & 4.6357(4)\{14\} \\
119m& 7/2 &0.621(7)\{12\}      & 4.6325(8)\{13\} \\
120 & 4   & 0.6538(11)\{133\} & 4.6361(1)\{14\} \\
120m1& 0   & 0.678(2)\{14\} & 4.6387(2)\{15\} \\
120m2&7   & 0.659(2)\{13\} & 4.6366(2)\{14\} \\
121 & 7/2 & 0.6997(14)\{144\} & 4.6410(2)\{15\} \\
121m& 1/2 & 0.727(2)\{15\} & 4.6440(2)\{16\} \\
\hline\hline
\end{tabular}
\end{table}

\begin{figure}
    \centering
    \includegraphics[width=0.45\textwidth]{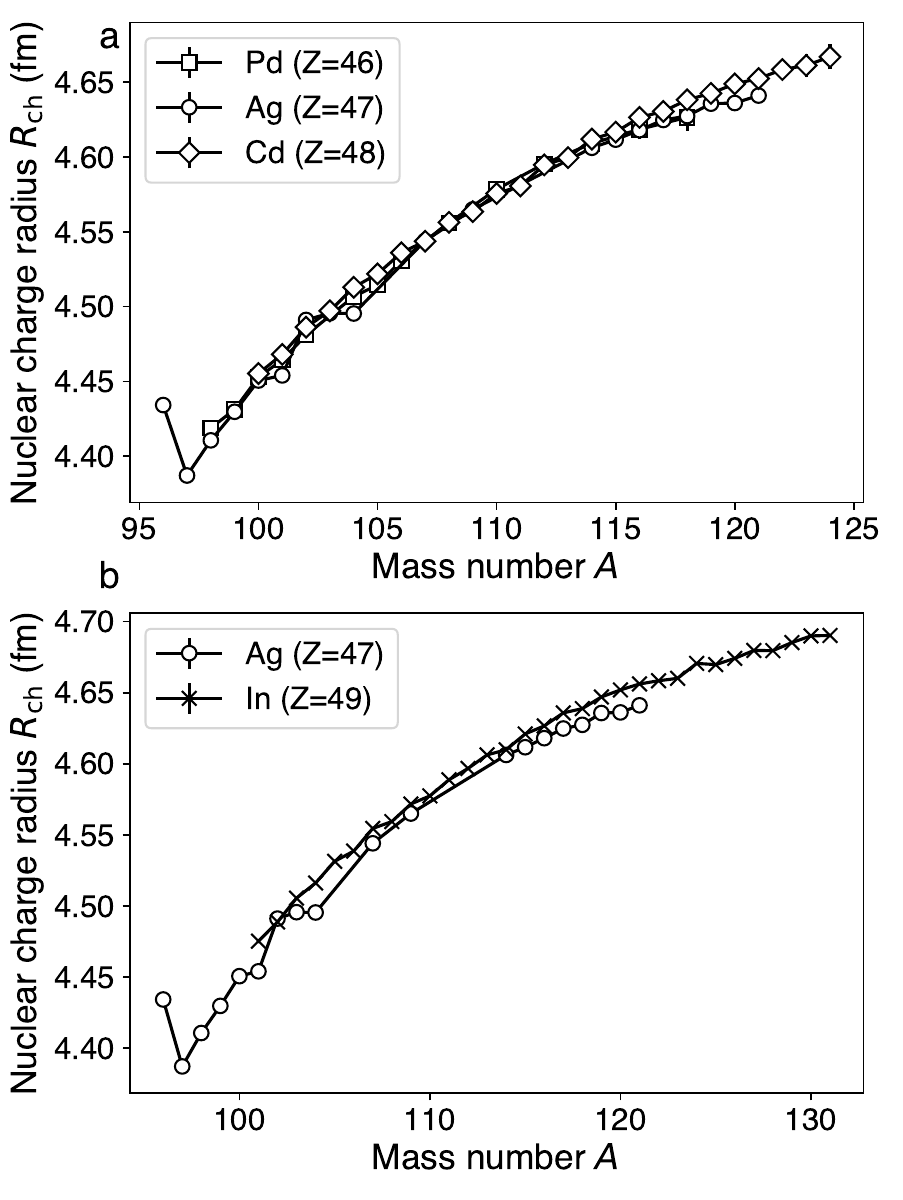}
    \caption{Comparison of absolute nuclear charge radii across the isotopic chain in the vicinity of Ag. {(a)} Comparison with immediately neighboring even-$Z$ elements Pd and Cd. {(b)} Comparison with the next odd-$Z$ element In. Values for Pd, Cd, and In are taken from Refs.~\cite{Geldhof2022,FrickeHeiligPd,Hammen2018,FrickeHeiligCd,Sahoo2020Indium}. The uncertainties for the Ag data points do not include the error on the absolute radius of the $^{109}$Ag reference isotope, and therefore the comparison on this plot should be considered qualitative.}
    \label{fig:chain_comparison}
\end{figure}

Using the charge radius for $^{109}$Ag recommended by Fricke and Heilig, the absolute charge radii for all Ag isotopes are derived and listed in Table~\ref{table:dr2}. A further comparison between $R_{\rm{ch}}$ in Ag and neighboring isotopic chains is made in Fig.~\ref{fig:chain_comparison}. The uncertainties for $R_{\rm{ch}}$ in Table~\ref{table:dr2} include only those propagated from the isotope-shift factors and experimental error bars and do not include the unknown common uncertainty of the absolute radius for $^{109}$Ag used as the reference value. The comparison with the even-proton Pd ($Z=46$) and Cd ($Z=48$) isotopic chains suggests only weak element dependence in this region. Further measurements of neutron-rich Ag and In isotopes via laser spectroscopy are necessary to show whether the difference in nuclear size on the neutron-rich side persists when approaching and going beyond $N=82$, and whether it is linked to differences in sphericity.

\section{Conclusion} \label{sec:conclusion}
In this work, high-accuracy \textit{ab initio} electronic structure calculations are performed with the relativistic coupled cluster method with single, double, triple, and perturbative quadruple excitation amplitudes, CCSDT(Q), for the lowest-lying electronic states of the neutral Ag atom. QED effects are shown to significantly contribute to the considered electronic properties, particularly to the mass shift factor, where their magnitude exceeds the achieved theoretical uncertainty. Isotope-shift factors and hyperfine constants are calculated for the electronic states relevant to recent laser spectroscopy experiments with radioactive isotopes. Literature NMR measurements of the magnetic dipole moment of the stable isotopes $^{107,109}$Ag are re-interpreted using a shielding constant calculation in this work. 

Nuclear magnetization distribution parameters are extracted for neutron-rich Ag isotopes, although the present uncertainties preclude firm nuclear-structure conclusions.

The isotope-shift factors and hyperfine constants calculated in this work can be used in the analysis of future laser spectroscopy experiments on radioactive Ag isotopes to extract accurate values of the nuclear moments and charge radii. Moreover, we call for further laser-spectroscopic experiments to measure the hyperfine structure of the 328-nm $5s\,^2S_{1/2} \rightarrow 5p\,^2P^{o}_{3/2}$ and 338-nm $5s\,^2S_{1/2} \rightarrow 5p\,^2P^{o}_{1/2}$ transitions in more Ag isotopes and with improved experimental precision, to extract more precise quadrupole moments (via the $^2P^{o}_{3/2}$ state), nuclear $g$-factors (via the $^2P^{o}_{1/2}$ state), and the BW parameter $B_s$, as done in this work. Such measurements will be instrumental in investigating the evolution in nuclear structure close to $Z=50$, $N=82$, and the relation between the nuclear $g$-factor and the extended nuclear magnetization distribution, as studied via the $B_s$ parameter.

Lastly, both the nuclear $g$-factor and the $B_s$ parameter can be considered as nuclear reference data, required to predict the hyperfine splitting in the Ag atom and its compounds in electronic structure calculations.

\section*{acknowledgments}
We thank Carlos Mario Fajardo Zambrano and Gerda Neyens for useful discussions.
Electronic structure calculations have been carried out using computing resources of the federal collective usage center Complex for Simulation and Data Processing for Mega-science Facilities at National Research Centre ``Kurchatov Institute'', \url{http://ckp.nrcki.ru/}. 
The work of L.V.S. at NRC ``Kurchatov Institute'' -- PNPI on electronic structure calculations was supported by the Russian Science Foundation under grant ~no.~26-12-00350, \url{https://rscf.ru/project/26-12-00350/}. The work of G.P. at Tsinghua University on electronic structure calculations was supported by the National Natural Science Foundation of China (No. 12341401 and No. 12274253). B.v.d.B., M.A.-K., and R.d.G. carried out the re-analysis of literature data and led the nuclear-structure discussions. All authors contributed to the manuscript writing and editing.

\section*{DATA AVAILABILITY}The data that support the findings of this study are available from the corresponding author upon reasonable request.


%


\appendix
\section{APPENDIX A: COMPUTATIONAL METHODS}\label{sec:methods}

\subsection{NMR shielding constant}
The magnetic dipole moments of $^{107}$Ag and $^{109}$Ag are typically used as reference values to extract the magnetic dipole moments of other isotopes using Eq.~\eqref{eq:opticalmu}. Their values have been measured using high-precision nuclear magnetic resonance (NMR) experiments of Ag in water and heavy water relative to the Larmor frequency of the proton and deuteron in the same solution~\cite{sahm1974}. These values were uncorrected for magnetic shielding effects.

In the current work, we re-evaluate the magnetic moments of $^{107}$Ag and $^{109}$Ag from Ref.~\cite{sahm1974} by calculating the shielding constant of the Ag(H$_2$O)$_4^+$ cation. This cation models the Ag$^+$ ions in water at infinite dilution. The geometry parameters of Ag(H$_2$O)$_4^+$ were optimized using the relativistic DFT method employing the PBE0~\cite{pbe0} functional and the uncontracted Dyall's AE3Z~\cite{Dyall:2007b,Dyall:12} basis set for Ag and the aug-cc-pVTZ-DK~\cite{Dunning:89,Kendall:92,augCCPVXZDK} basis set for light atoms. The optimized Ag--O bond length value was obtained at 2.384~\AA.

The diamagnetic component of the shielding constant was computed using the relativistic linear-response DFT method~\cite{olejniczak2012,ilias2013,aucar1999,dirac19} with the PBE0 functional. For this calculation, we employed the uncontracted AE4Z~\cite{Dyall:07,Dyall:12} basis set for all atoms involved. The paramagnetic part of $\sigma$ was calculated using the relativistic coupled cluster method with single, double, and perturbative triple excitations, CCSD(T)~\cite{Visscher:96a}, employing the finite field approach as suggested in Ref.~\cite{skripnikov2018}. Due to the high computational cost associated with this method, the calculation was performed using the uncontracted all-electron triple-zeta AE3Z basis set for Ag and the AE2Z~\cite{Dyall:2007b} basis set for light atoms. 
To address basis set incompleteness, we also computed the basis-set correction for the paramagnetic component within the DFT framework. This correction was calculated as the difference between results obtained using the uncontracted all-electron quadruple-zeta AE4Z basis set for all atoms and the same basis set employed in the CC calculations. Finally, we estimated the contribution of the Gaunt electron-electron interaction within the DFT approach and accounted for the solvent effect calculated in Ref.~\cite{antusek2020}.

\subsection{Atomic hyperfine constants and excitation energies}
To account for the leading relativistic and electronic correlation effects in a balanced way, we performed calculations using the relativistic CCSD(T) method~\cite{Visscher:96a,Bartlett:2007}, employing the Dirac-Coulomb Hamiltonian. This correlation calculation included all electrons of the Ag atom without any energy cutoff for virtual orbitals, which is essential for accurately considering the correlation contributions of the inner-core electrons~\cite{Skripnikov:17a,Skripnikov:15b}.
One-electron functions (atomic orbitals) were obtained using the Dirac–Hartree–Fock method for the Ag$^+$ ion and a Gaussian nuclear charge model~\cite{Visscher:1997}.
We employed Dyall's uncontracted augmented all-electron quadruple-zeta AE4Z basis set~\cite{Dyall:07,Dyall:12}, augmented with additional diffuse and tight Gaussian functions. Additionally, we replaced $g$-, $h$-, and $i$-type functions with uncontracted natural-like functions~\cite{Skripnikov2020BW,Skripnikov:13a}. 
To obtain these functions, we performed scalar-relativistic all-electron CCSD(T) calculations of the Ag atom and derived effective one-particle reduced density matrices for the $5s~^2S$ and $5p~^2P^{o}$ electronic states of Ag.

These calculations were performed using a very large even-tempered basis set that included, in particular, 15$g$-, 15$h$-, and 15$i$-type functions. Practically, it was not possible to perform such calculations within the four- or two-component wave function frameworks. In the scalar-relativistic correlation calculation, one obtains density matrices with elements $D^{\alpha/\beta}_{\mu,\nu}$ for two spin projections, $\alpha$ and $\beta$, in the representation of the scalar atomic basis functions $\chi_\mu$. In the procedure used, we averaged over the spin projections, atomic-basis-function spatial projections, and all considered electronic states. This procedure yielded an effective matrix of the one-electron electronic-state-averaged density operator for an atom in the atomic orbital basis set representation. The eigenfunctions with corresponding eigenvalues of this operator were then calculated. In principle, functions with the highest eigenvalues (above some threshold) can be used as a natural-like basis set in a contracted form for further relativistic calculations~\cite{Skripnikov:15a,skripnikov2015TaN,Skripnikov:16b,AthanasakisKaklamanakis:2025}. However, such a contracted form of basis functions cannot currently be used in relativistic four-component calculations. Therefore, following Ref.~\cite{Skripnikov2020BW}, the obtained contracted basis functions were re-expanded into a much smaller number of optimized primitive basis functions.

In the optimization procedure, we took into account natural occupations to ensure that the most important basis functions were described accurately. This method allowed us to obtain a compact set of primitive Gaussian basis functions, which is especially important for the high harmonics due to the high computational cost of including each such function in the relativistic four-component correlation calculation. In total, the ``reference'' basis set, referred to as LBas, consisted of 37$s$-, 29$p$-, 26$d$-, 16$f$-, 8$g$-, 5$h$-, and 3$i$-type Gaussian functions, and can be succinctly written as [37,29,26,16,8,5,3]. The hyperfine structure constants were calculated using the finite-field approach.

Next, we calculated the basis-set correction in three steps. 
In the first step, we replaced all $s$-, $p$-, $d$-, and $f$-type functions with an even-tempered series, resulting in the [50,50,44,24,8,5,3] basis set. 
In the second step, we increased the number of $g$-, $h$-, and $i$-type functions relative to the reference basis set, resulting in the [37,29,26,16,10,7,5] basis set. 
Finally, we accounted for the contribution of $k$-type functions using the [37,29,26,16,8,5,3,3] basis set. Each basis-set correction was calculated as the difference between the relativistic CCSD(T) values obtained with the increased basis set and those obtained with the reference basis set. All electrons were correlated, and the virtual orbitals cutoff was set to 10,000 Hartree.
The contribution of $k$‑type functions was negligible for all calculated quantities except the excitation energy (EE) of the $4d^95s^2\,^2$D$_{5/2}$ state. Therefore, we also estimated the extrapolation correction for higher harmonics using the same approach as in Refs.~\cite{Skripnikov:2021a,AthanasakisKaklamanakis:2025}.

In the next stage of the calculation, we increased the level of treatment of electron correlation effects. To achieve this, we calculated the following corrections: (i) first, we increased the level of electron correlation treatment from CCSD(T) to CCSDT-3~\cite{Noga:CCSDT-3:87}. Unlike CCSD(T), the CCSDT-3 approach is iterative but still lacks some diagrams of the full iterative CC model with single, double, and triple excitation amplitudes (CCSDT)~\cite{Bartlett:2007}. For this calculation, we used the MBas basis set, based on Dyall's uncontracted all-electron AE3Z basis set~\cite{Dyall:07,Dyall:12}, augmented with additional diffuse and tight Gaussian functions. In total, this basis set consists of [32,25,23,13,3] basis functions. In these calculations, all electrons were correlated, and the cutoff for virtual orbital energies was set to 10,000 Hartree. (ii) To account for the effect of full triple excitation amplitudes, we calculated the difference between the results obtained using the CCSDT and the CCSDT-3 methods. Due to the high computational cost of the CCSDT method, 28 inner-core electrons were excluded from the correlation calculations, and the cutoff for virtual energies was set to 50 Hartree. (iii) Finally, we estimated the contribution of quadruple excitations as the difference between results obtained using the CCSDT(Q)~\cite{Kallay:6} method (with single, double, triple, and perturbative quadruple excitations) and the CCSDT method. The same computational parameters as in the previous correction were employed for the calculation of hyperfine constants, while for the calculation of the EEs, the cutoff for virtual energies was set to 10 Hartree.

Finally, we enhanced the electronic Hamiltonian by incorporating additional interactions beyond the Dirac-Coulomb one. Firstly, we calculated the contribution of the Breit electron-electron interaction. This was determined as the difference between the CCSD(T) results obtained with and without the Breit interaction term in the Hamiltonian using the SBas basis set [25,20,15,10,3], which is an approximation to the MBas set. 
Next, we estimated the order of magnitude of the QED effects' contribution to hyperfine constants by comparing the CCSD(T) results obtained using the LBas basis set with and without the inclusion of the effective VP and model SE Hamiltonians~\cite{Shabaev:13,Malyshev:2022}, using the implementation described in Ref.~\cite{Skripnikov:2021a}. It is important to note that the employed model QED approach~\cite{Shabaev:13} was designed to compute the QED contributions to the energies~\cite{Shabaev:13,Malyshev:2022} and does not include the complete set of diagrams necessary to account for the SE (and VP) effect contributions to the hyperfine constants. This limitation is reflected in the corresponding uncertainty estimation.

\subsection{Atomic isotope-shift factors}

To calculate the IS factors, we followed the outlined procedure with several modifications described below, motivated by the balance between target uncertainty and required computational resources. Firstly, to estimate the QED contributions to the FS and MS constants, we applied recently proposed approaches~\cite{skripnikov2024isotopeQED,Anisimova:2022}, as briefly described in the main text.

By default, all calculations employed the Gaussian nuclear charge distribution model. For the FS constant, we calculated the correction due to the choice of nuclear charge distribution model as the difference between the $F$ constant values obtained using Fermi and Gaussian nuclear charge distributions, referred to in this work as the ``Nuc. model'' correction.

Next, for the ``CCSDT-3 $-$ CCSD(T)'' correction to the MS and FS constants, we used the SBas basis set [25,20,15,10,3] instead of the MBas one (see the corresponding uncertainty estimation consequences due to this modification below). 

The SBas basis set was also employed to calculate the ``CCSDT $-$ CCSDT-3'' and ``CCSDT(Q) $-$ CCSDT'' corrections to the MS factors. As for the corresponding hyperfine structure and EEs calculations described above, 28 inner-core electrons were excluded from the correlation treatment, and the cutoff energy for virtual orbitals included in correlation calculation was set to 50 Hartree. To account for the contribution of correlation effects beyond CCSDT-3 for the FS, we used the MBas basis set but reduced the cutoff energy to 10 Hartree.

For the reference CCSD(T) calculation of the NMS and FS factors, we employed the cutoff for virtual orbitals to 10,000 Hartree, and still used the LBas basis set as in the case of EEs and hyperfine constants. Due to technical reasons, the reference CCSD(T) calculation for the SMS factors was performed using the MBas basis set. However, an additional basis-set correction is included, which is computed as the difference between the CCSD results obtained using the LBas and MBas basis sets. 
Finally, due to the very small values of basis-set corrections obtained in the calculation of EEs and hyperfine constants (see below), we did not apply the corresponding correction for the IS constants. However, we made a corresponding uncertainty estimation due to basis set incompleteness.

Electronic structure calculations were performed with the {\sc dirac}~\cite{dirac19,Saue:2020}, {\sc mrcc}~\cite{kallay2020,Kallay:1,Kallay:2}, and {\sc exp-t}~\cite{EXPT_website,Oleynichenko_EXPT,Zaitsevskii:2023} codes. To generate a compact set of basis functions with high angular momenta, we used the {\sc natbas} code developed in Refs.~\cite{Skripnikov2020BW,Skripnikov:13a} and the {\sc cfour}~\cite{Matthews:CFOUR:20} code to generate effective one-particle density matrices at the CCSD(T) level. The code for calculating matrix elements of considered operators over atomic bispinors was developed in Refs.~\cite{Skripnikov:16b,Schmidt:2018}. We also used the code developed in Ref.~\cite{Skripnikov:2021a} to incorporate SE and VP QED effects for EEs and estimate them for hyperfine constants. We used the code developed in Ref.~\cite{skripnikov2024isotopeQED} for QED contributions to FS and MS factors. To treat the nuclear Fermi charge distribution, we used the code developed in Ref.~\cite{Skripnikov:2024a}. Matrix elements of the NMS and SMS operators were calculated using the code developed in Ref.~\cite{Penyazkov:2023}. Two-electron integrals of the Breit interaction operator over atomic bispinors were calculated using the code developed in Ref.~\cite{Maison:2019}.

\subsection{Uncertainty estimation}

The following procedure was employed to estimate uncertainties in electronic structure calculations:

(i) For the hyperfine constants and EEs, the ``CCSDT-3 $-$ CCSD(T)'' correction was computed using MBas and SBas basis sets. The uncertainty of this correction was determined as the difference between the values obtained with these two basis sets. For the nuclear recoil IS constants, this correction was also performed using two basis sets, but only for the $5s ^2S_{1/2} \rightarrow 5p~^2P^{o}_{1/2}$ transition. The deviation between the two corresponding values was found to be within 10\%. For other transitions, this correction was calculated solely within the SBas basis set. We set the uncertainty of this correction to 30\% of its value. For the FS constant, this correction was calculated exclusively within the SBas basis set, and we set the corresponding uncertainty to 100\% of this correction.

(ii) As a measure of the uncertainty of the ``CCSDT $-$ CCSDT-3'' correction for the hyperfine constants (except for $\dbtilde{A}_{\rm BW,el}$) and EEs, we compared its values in the MBas and SBas basis sets. For IS atomic factors and $\dbtilde{A}_{\rm BW,el}$, the uncertainty of this correction was set to 100\% of its value.

(iii) The uncertainty of the ``CCSDT(Q) $-$ CCSDT'' correction was set to 100\% of its value for all computed parameters, indicating that this correction can be considered a measure of correlation effects beyond the CCSDT(Q) model.

(iv) The uncertainty due to the nuclear charge distribution model correction for the FS constant was set to 50\% of its value, reflecting the significant difference between the Gaussian and Fermi distribution models.

(v) The uncertainty of the basis-set correction for EEs and hyperfine constants was set to 100\% of its value. 
This is quite a conservative estimate, as this correction implies a significant modification (e.g., completely replacing all $s$-, $p$-, $d$-, and $f$‑type functions with an even‑tempered series) and increase in the basis set size (see above). 
Because the basis-set corrections to $\Tilde{A}_0$ are small, the corresponding corrections to $\dbtilde{A}_{\rm BW,el}$ were neglected. Nevertheless, the associated uncertainties were estimated by multiplying the CCSD(T) values of $\dbtilde{A}_{\rm BW,el}$ by the relative basis-set corrections to $\Tilde{A}_0$.
For the IS constants, the uncertainty due to basis set incompleteness was calculated as the difference between the values of these constants obtained using the LBas and MBas basis sets at the CCSD(T) level of theory for NMS and FS constants and the CCSD level for the SMS constants.

(vi) The uncertainty of the Breit interaction correction was set to 50\%, indicating the lack of the $\omega$-dependent term.

(vii) The uncertainty of the QED correction was set to 100\% of its value for hyperfine constants due to its very approximate application in the present case. For EEs and IS constants, we used specifically designed model QED operators of EEs~\cite{Shabaev:13,Skripnikov:2021a}, FS~\cite{skripnikov2024isotopeQED}, and MS~\cite{Anisimova:2022,skripnikov2024isotopeQED} constants. Therefore, the corresponding uncertainties were set to 30\% of their values.

The final uncertainty was calculated as the square root of the sum of the squares of all the uncertainties described above.

\section{APPENDIX B: EXTRACTED MAGNETIC DIPOLE MOMENTS}
Table~\ref{tab:full_dipole_moments} lists the measured hyperfine $A$-factors from literature optical experiments used to extract the magnetic dipole moments and nuclear magnetization distribution parameters in Table~\ref{tab:new_dipole_moments}. For $^{107,109, 113-121}$Ag, where the magnetic dipole moments can be determined using more than one approach, the preferred moments are typeset in bold font.

\begin{table*}[]
\caption{Summary of the measured hyperfine $A$-factors, uncorrected magnetic dipole moment ($\mu_{\rm{uncorr}}$) from Eq.~\eqref{eq:opticalmu_approx} using $\mu(^{109}\rm{Ag)}$ from Eq.~\eqref{mu109}, corrected magnetic dipole moment using the Moskowitz-Lombardi rule~\cite{moskowitz1973} ($\mu_{\rm{ML}}$), corrected magnetic dipole moment using Eq.~\eqref{Aparam2} ($\mu^{\rm{1-eq}}_{\rm{CC}}$), magnetic dipole moment from re-evaluated NMR ($\mu_{\text{re-eval NMR}}$), corrected magnetic dipole moment using Eqs.~\eqref{eq:2x2_1},\eqref{eq:2x2_2} ($\mu^{\rm{2-eq}}_{\rm{CC}}$) and the absolute hydrogen-like Bohr-Weisskopf nuclear parameter ($B_s$) for all optically measured Ag isotopes. Uncertainties originating from experimental and theoretical contributions are given in parentheses and curly brackets, respectively. The spin assignments are from the ENSDF database~\cite{ENSDF}, unless specified otherwise. For $\mu_{\rm{uncorr}}$, $^{109}$Ag is the reference isotope. The preferred dipole moments are given in bold. For $^{98}$Ag, two different potential spin assignments are considered.}\label{tab:full_dipole_moments}
\resizebox{\textwidth}{!}{
\begin{threeparttable}
\begin{tabular}{ccccccccc}
\hline
\hline
\hspace{0.2cm}Isotope\hspace{0.2cm}    & \hspace{0.2cm}$I$\hspace{0.2cm}   & $A(^2S_{1/2})$ (MHz) &   $\mu_{\text{uncorr}}$ ($\mu_N$) & & \hspace{0.2cm}$\mu_{\rm{ML}}$ ($\mu_N$)\hspace{0.2cm}      &     \\\hline
97   & (9/2)       &  +10600(200)~\cite{ferrer2014}  &  +6.30(11)  &       &  \textbf{+6.12(12)}   &  \\
98   & (6)         &  +6020(90)~\cite{ferrer2014}  & +4.77(7) &  & \textbf{+4.64(7)} &  \\
     & (5)        &  +7120(110)~\cite{ferrer2014}  & +4.70(7) &  & \textbf{+4.57(7)} &  \\
99   & (9/2)      &  +10050(50)~\cite{ferrer2014}  &  +5.97(3)   &       &   \textbf{+5.80(3)}  &  \\
100  & (5)         &  +6810(40)~\cite{ferrer2014}  & +4.50(3) &  & \textbf{+4.37(3)} &  \\
\midrule
\hspace{0.2cm}Isotope\hspace{0.2cm} & \hspace{0.2cm}$I$\hspace{0.2cm} & $A(^2D_{5/2})$ (MHz) & $\mu_{\rm{uncorr}}$ ($\mu_N$) & & & \hspace{0.2cm}$\mu^{\text{1-eq}}_{\rm{CC}}$ ($\mu_N$)\hspace{0.2cm} \\\midrule
101 & 9/2 &  +694.4(14)~\cite{dinger1989} & \textbf{+5.621(11)} &  &     & +5.606(11)\{18\} \\
103 & 7/2 &  +703.2(14)~\cite{dinger1989} & \textbf{+4.427(9)} &  &     & +4.417(9)\{14\}   \\
104 & 5   &  +435.3(6)~\cite{dinger1989} & \textbf{+3.915(5)} &  & & +3.906(5)\{12\}   \\
105m & 7/2  &  +700.4(24)~\cite{dinger1989} & \textbf{+4.409(15)} & &      & +4.400(15)\{14\}  \\
\midrule
\hspace{0.2cm}Isotope\hspace{0.2cm}    & \hspace{0.2cm}$I$\hspace{0.2cm}   & \hspace{0.2cm}$A(^2S_{1/2})$ (MHz)\hspace{0.2cm} & $\mu_{\text{uncorr}}$ ($\mu_N$)  & \hspace{0.2cm}$A(^2P_{3/2})$ (MHz)\hspace{0.2cm} & \hspace{0.2cm}$\mu_{\text{re-eval NMR}}$ ($\mu_N$)\hspace{0.2cm} & \hspace{0.2cm}$\mu^{\text{2-eq}}_{\rm{CC}}$ ($\mu_N$)\hspace{0.2cm}    & \hspace{0.2cm}$B_s$ ($10^5$\,MHz/$\mu_N$)\hspace{0.25cm}   \\\midrule

107  & 1/2 & $-$1712.512111(18)~\cite{dahmen1967}     & $-$0.113089(9) &  $-$31.7(5)~\cite{Carlsson1990} & \textbf{$-$0.113558(9)}  & $-$0.1128(20)\{4\}  & +5.00(1)\{46\} \\
109  & 1/2 & $-$1976.932075(17)~\cite{dahmen1967}     &    --          &  $-$36.7(7)~\cite{Carlsson1990} & \textbf{$-$0.130551(10)} & $-$0.1307(28)\{4\} & +4.39(1)\{46\} \\
\midrule
\hspace{0.2cm}Isotope\hspace{0.2cm}    & \hspace{0.2cm}$I$\hspace{0.2cm}   & \hspace{0.2cm}$A(^2S_{1/2})$ (MHz)\hspace{0.2cm} & $\mu_{\rm{uncorr}}$ ($\mu_N$)  & \hspace{0.2cm}$A(^2P_{3/2})$ (MHz)\hspace{0.2cm} & \hspace{0.2cm}$\mu_{\rm{ML}}$ ($\mu_N$)\hspace{0.2cm} & \hspace{0.2cm}$\mu^{\text{2-eq}}_{\rm{CC}}$ ($\mu_N$)\hspace{0.2cm}    & \hspace{0.2cm}$B_s$ ($10^5$\,MHz/$\mu_N$)\hspace{0.25cm}   \\\midrule
113m & 7/2 & +9609(5)~\cite{degroote2024AgPLB} & +4.442(2)   & +173.7(13)~\cite{degroote2024AgPLB,vandenborne2025} & +4.315(2)   & \textbf{+4.316(36)\{14\}} & +0.1(13)\{7\} \\
114 & 1 & +19215(22)~\cite{vandenborne2025}              & +2.538(3)   & +352(4)~\cite{vandenborne2025}    & +2.464(3)   & \textbf{+2.503(32)\{8\}} & +2.3(19)\{7\}   \\
115m & 7/2 & +9556(2)~\cite{degroote2024AgPLB} & +4.4174(10) & +173.5(7)~\cite{degroote2024AgPLB,vandenborne2025}  & +4.2916(12) & \textbf{+4.313(19)\{14\}} & +0.8(7)\{7\}  \\
116m1 & 4$^\bullet$ & +5329(5)~\cite{vandenborne2025}    & +2.815(3)   & +96.6(10)~\cite{vandenborne2025}  & +2.734(3)   & \textbf{+2.744(32)\{9\}} & +0.5(18)\{7\}   \\
116m2 & 7$^\bullet$ & +3304.0(5)~\cite{vandenborne2025}  & +3.0546(5)  & +60.10(12)~\cite{vandenborne2025} & +2.9665(7)  & \textbf{+2.989(7)\{10\}} & +1.1(3)\{7\}    \\
117m & 7/2$^\bullet$ & +9486(2)~\cite{degroote2024AgPLB} & +4.3850(10) & +172.5(4)~\cite{degroote2024AgPLB,vandenborne2025}  & +4.2601(12) & \textbf{+4.289(11)\{14\}} & +1.1(4)\{7\}  \\
118 & 4$^\bullet$ & +5911(2)~\cite{vandenborne2025}      & +3.123(11)  & +106.0(12)~\cite{vandenborne2025} & +3.0328(12) & \textbf{+3.007(38)\{10\}} & $-$1.3(19)\{7\} \\
118m2 & 7$^\bullet$ & +3812.9(10)~\cite{vandenborne2025} & +3.5251(10) & +68.8(4)~\cite{vandenborne2025}   & +3.4240(11) & \textbf{+3.418(22)\{11\}} & $-$0.2(10)\{7\} \\
119m & 7/2$^\bullet$ & +9581(2)~\cite{degroote2024AgPLB} & +4.4289(10) & +177.2(5)~\cite{degroote2024AgPLB,vandenborne2025}  & +4.3028(12) & \textbf{+4.415(14)\{14\}$^\star$} & +3.9(5)\{7\}$^\star$  \\
120 & 4$^\bullet$ & +6045(4)~\cite{vandenborne2025}      & +3.194(2)   & +107.9(16)~\cite{vandenborne2025} & +3.102(2)   & \textbf{+3.059(51)\{10\}} & $-$2.1(26)\{7\} \\
120m2 & 7$^\bullet$ & +3651(3)~\cite{vandenborne2025}    & +3.375(3)   & +65.5(9)~\cite{vandenborne2025}   & +3.278(3)   & \textbf{+3.252(50)\{10\}} & $-$1.3(24)\{7\} \\
121  & 7/2$^\bullet$ & +9610(3)~\cite{degroote2024AgPLB} & +4.4423(14) & +172.5(16)~\cite{degroote2024AgPLB,vandenborne2025} & +4.316(2)   & \textbf{+4.283(44)\{14\}} & $-$1.1(16)\{7\}    \\
\hline
\hline
\end{tabular}
\begin{tablenotes}
    \item[$(\bullet)$] Spin assignment from Refs.~\cite{degroote2024AgPLB,vandenborne2025}.
    \item[$(\star)$] Possible outlier (see text for details).
\end{tablenotes}
\end{threeparttable}
}
\end{table*}

\end{document}